\magnification=1200
\hoffset=.0cm
\voffset=.0cm
\baselineskip=.55cm plus .55mm minus .55mm

%
%
\def\ref#1{\lbrack#1\rbrack}
%
%
%
%
\input amssym.def
\input amssym.tex
%
%
\font\teneusm=eusm10                    
\font\seveneusm=eusm7                   
\font\fiveeusm=eusm5                 
%
%

%
%
\font\cps=cmcsc10
%
%
\newfam\eusmfam
\textfont\eusmfam=\teneusm
\scriptfont\eusmfam=\seveneusm
\scriptscriptfont\eusmfam=\fiveeusm

\def\proclaim #1. #2\par{\medbreak{\cps #1.\enspace}{\it #2}\par\medbreak}
%
%
%
%
\def\sgn{\hskip 1pt{\rm sgn}\hskip 1pt}
\def\dtr{\hskip 1pt{\rm det}\hskip 1pt}

\def\ker{\hskip 1pt{\rm ker}\hskip 1pt}

\def\tr{\hskip 1pt{\rm tr}\hskip 1pt}
\def\Tr{\hskip 1pt{\rm Tr}\hskip 1pt}
\def\ad{\hskip 1pt{\rm ad}\hskip 1pt}
\def\Ad{\hskip 1pt{\rm Ad}\hskip 1pt}
\def\Lie{\hskip 1pt{\rm Lie}\hskip 1pt}

\def\Gau{\hskip 1pt{\rm Gau}\hskip 1pt}

\def\Conn{\hskip 1pt{\rm Conn}\hskip 1pt}

\def\hst1{\hskip 1pt}

%
%
%
%

\hrule\vskip.5cm
\hbox to 16.5 truecm{May 1996  \hfil DFUB 96--15}
\hbox to 16.5 truecm{Version 1  \hfil hep-th/9607003}
\vskip.5cm\hrule
\vskip.9cm
\centerline{\bf REDUCIBILITY AND GRIBOV PROBLEM IN}   
\centerline{\bf TOPOLOGICAL QUANTUM FIELD THEORY}   
\vskip.4cm
\centerline{by}
\vskip.4cm
\centerline{\bf Roberto Zucchini}
\centerline{\it Dipartimento di Fisica, Universit\`a degli Studi di Bologna}
\centerline{\it V. Irnerio 46, I-40126 Bologna, Italy}
\vskip.9cm
\hrule
\vskip.6cm
\centerline{\bf Abstract} 
\vskip.4cm
\par\noindent
In spite of its simplicity and beauty, the Mathai-Quillen formulation of 
cohomological topological quantum field theory with gauge symmetry 
suffers two basic problems: $i$) the existence of reducible field 
configurations on which the action of the gauge group is not free and 
$ii$) the Gribov ambiguity associated with gauge fixing, i. e. the lack 
of global definition on the space of gauge orbits of gauge fixed 
functional integrals. In this paper, we show that such problems are in fact 
related and we propose a general completely geometrical recipe for their 
treatment. The space of field configurations is augmented in such a way to 
render the action of the gauge group free and localization is suitably 
modified. In this way, the standard Mathai--Quillen formalism can be 
rigorously applied. The resulting topological action contains the ordinary 
action as a subsector and can be shown to yield a local quantum field theory,
which is argued to be renormalizable as well. The salient feature of our 
method is that the Gribov problem is inherent in localization, and thus can 
be dealt with in a completely equivariant setting, whereas gauge fixing is 
free of Gribov ambiguities. For the stratum of irreducible gauge orbits, 
the case of main interest in applications, the Gribov problem is solvable. 
Conversely, for the the strata of reducible gauge orbits, the Gribov 
problem cannot be solved in general and the obstruction may be described
in the language of  sheaf theory. The formalism is applied to the 
Donaldson--Witten model.
\vfill\eject

\vskip .4cm
{\bf 0. Introduction}
\vskip .4cm
\par
Topological quantum field theories are complicated often
fully interacting field 
theories having no physical degrees of freedom. Yet they can be solved 
exactly and the solution is rather non trivial. Expectation values of 
topological observables provide topological invariants of the manifolds
on which the fields propagate. These invariants are independent from the 
couplings and to a large extent from the interactions between the fields. 
At the same time topological field theories are often topological sectors 
of ordinary field theories. In this way they are convenient testing grounds
for subtle non perturbative 
field theoretic phenomena such as duality. See refs. \ref{1--10} 
for comprehensive reviews on the subject and complete referencing.

As well-known, there are two basic kinds of topological field theories:
those of Schwarz type and those of cohomological type. The prototype of the 
Schwarz type topological field theories is the Chern--Simon model \ref{11}.
To the second group there belong the Donaldson--Witten model \ref{12}, the 
topological sigma model \ref{13} and topological two--dimensional 
gravity \ref{14}. In this paper, we shall concentrate on topological field 
theories of cohomological type.

A great impetus to the development of cohomological topological 
quantum field theory
has come from the realization that they may be understood in the framework
of equivariant cohomology of infinite dimensional vector bundles \ref{1,
15--19}
and realized as Mathai-Quillen integral representations of Euler classes 
\ref{20--23}. This has lead 
to a clear geometric interpretation of such field theoretic models 
providing a rather general framework for their understanding and has 
furnished a simple tool kit for the construction of other models. The 
formalism has been extended also to gauge theories in a way that
respects general principles of field theory such as locality and 
renormalizability.

In spite of its simplicity and beauty, the Mathai-Quillen formulation of 
cohomological topological quantum field theory with gauge symmetry 
suffers two basic problems. The first consists in the existence of reducible 
field configurations on which the action of the gauge group is not free.
Various attempts at its solution in various model have appeared \ref{24--27}. 
The second is the Gribov ambiguity associated with gauge fixing, i. e. 
the lack of global definition on the space of gauge orbits of gauge fixed 
functional integrals. It has been tackled in ref. \ref{28--30} in topological 
two--dimensional gravity.

In this paper, we propose a general recipe for the treatment of the problems 
just mentioned and explore its consequences. The space of field configurations 
$\cal A$ carrying the non free right action of a group $\widehat{\cal G}$ 
is substituted by the larger space ${\cal P}={\cal N}\times{\cal A}\times
{\cal G}$, where $\cal N$ is a stratum of the $\widehat{\cal G}$ 
orbit space of $\cal A$ and $\cal G$ is the subgroup of 
$\widehat{\cal G}$ of the elements path connected to the identity.
$\cal G$ acts freely to the right
on $\cal P$ in natural fashion. In this way, the standard 
Mathai--Quillen formalism can be rigorously applied to $\cal P$. 
The Mathai--Quillen localization sector is then suitably augmented 
to eliminate the extra degrees of freedom associated with the factor 
$\cal A$ of the $\cal G$ orbit space of $\cal P$, ${\cal N}\times{\cal A}$. 
A local topological action, which contains the customary action as 
a subsector and is argued to be renormalizable, is produced in this way. 

The topological quantum field 
theory yields a map from the equivariant cohomology of $\cal A$ 
to the cohomology of each element of an open covering $\{U_\alpha\}$ of 
$\cal N$. The Gribov problem is solved if, for a given equivariant
cohomology class $\cal O$ of $\cal A$, the corresponding cohomology class of 
each $U_\alpha$ is the restriction on $U_\alpha$ of a unique cohomology class 
$\varphi_{\cal O}$ of $\cal N$ depending only on $\cal O$. The salient 
feature of our method is that the Gribov problem is inherent in localization, 
and thus can be dealt with in a completely equivariant setting, whereas gauge 
fixing is free of Gribov ambiguities.
If $\cal N$ is the stratum of $\widehat{\cal G}$ regular irreducible orbits, 
the case of main interest in applications, the Gribov problem is 
shown to be solvable under certain rather general assumptions. 
If $\cal N$ is instead a stratum of $\cal G$ singular 
reducible orbits, then the Gribov problem cannot be solved in 
general and the obstruction may be characterized in a suitable sheaf 
theoretic framework. {\it This shows that reducibility and Gribov 
ambiguity are related aspects of topological quantum field theory}.

The formalism is applied to the Donaldson--Witten model as an illustration. 
It can also be applied to two--dimensional topological Yang--Mills and to 
topological QCD. In principle, it should work also for two--dimensional
topological gravity and topological string theory, though in these latter 
cases, a reexamination of gauge fixing is necessary.
\vskip .4cm
{\bf 1. Cohomological Topological Quantum Field Theory}
\vskip .4cm
\par
The field content of a topological quantum field theory of cohomological type 
is organized according to a certain algebraic structure called an operation. 
Recall that an operation is a quintuplet
$(\goth g,Z,j,l,s)$, where $\goth g$ is a Lie algebra, $Z$ is a graded 
associative algebra and $j(\xi)$, $l(\xi)$, $\xi\in\goth g$, and $s$ are 
graded derivations on $Z$ of degree $-1$, $0$, $+1$, respectively, satisfying 
Cartan's algebra:
$$
\eqalign{&[j(\xi),j(\eta)]\cr 
         &[l(\xi),l(\eta)]\cr 
         &[s,l(\eta)]\cr}
\eqalign{&\!\!=0, \cr 
         &\!\!=l([\xi,\eta]), \cr 
         &\!\!=0, \cr}
\qquad\qquad
\eqalign{&[l(\xi),j(\eta)]\cr 
         &[s,j(\eta)]\cr 
         &[s,s]\cr}
\eqalign{&\!\!=j([\xi,\eta]), \cr 
         &\!\!=l(\eta), \cr 
         &\!\!=0, \cr}
\eqno(1.1)
$$
where the above are graded commutators \ref{31}. Since $s^2=0$, one can 
define the cohomology of the differential complex $(Z,s)$. This is called 
ordinary $s$ cohomology. More importantly, one may consider 
the differential complex $(Z_{\rm basic},s)$, where $Z_{\rm basic}$ is the 
$s$ invariant subalgebra of $Z$ annihilated by all $j(\xi)$, $l(\xi)$, 
$\xi\in\goth g$. The corresponding cohomology is referred to
as basic $s$ cohomology. 

The main ingredients entering the construction of any cohomological
topological quantum field theory are the following.
\par\noindent
\item{$i$)} A Lie group $\cal G$ with Lie algebra $\Lie{\cal G}$.
\item{$ii$)} A principal $\cal G$ bundle $\pi_{\cal P}:\cal P\to\cal M$.
\item{$iii$)} A vector space $\cal U$ with a left $\cal G$ action.
\par\noindent
Using these, one can construct the relevant operation 
$(\Lie{\cal G},{\cal W},j,l,s)$. Here, ${\cal W}$ is the graded tensor 
algebra
$$
{\cal W}=W_1(\Lie{\cal G})\hat\otimes\Omega^*({\cal P})\hat\otimes 
W_{-1}({\cal U}^\vee)\hat\otimes W_{-2}(\Lie{\cal G})\hat\otimes 
W_{-1}(\Lie{\cal G})
\eqno(1.2)
$$
\footnote{}{}
\footnote{${}^1$}{For any vector space $V$, $W_p(V)=S(V^\vee)\otimes 
A(V^\vee)$, where $S(V^\vee)$ and $A(V^\vee)$ are the symmetric and 
antisymmetric algebras of $V^\vee$, respectively. 
$S^1(V^\vee)\simeq V^\vee$, $A^1(V^\vee)\simeq V^\vee$ carry 
degree $p$, $p+1$, for $p$ even, and $p+1$, $p$, for $p$ odd, respectively. 
For any manifold $X$, $\Omega^*(X)$ is essentially the exterior 
algebra of $X$. $\hat\otimes$ denotes graded tensor product.
See ref.  \ref{9}.}. 
The generators of the tensor factors in the given 
order are
$$
\eqalign{&\omega,~\Omega \cr 
         &p,~\phi \cr 
         &\rho,~\pi \cr 
         &\bar\Omega,~\bar\psi \cr
         &\bar\lambda,~\bar\chi \cr}
\qquad\qquad
\eqalign{&\deg\omega \cr 
         &\deg p \cr 
         &\deg\rho \cr
         &\deg\bar\Omega\cr
         &\deg\bar\lambda \cr}
\eqalign{&\!\!=\phantom{-}1,~ \cr 
         &\!\!=\phantom{-}0,~ \cr 
         &\!\!=-1,~ \cr
         &\!\!=-2,~ \cr
         &\!\!=-1,~ \cr}
\eqalign{&\deg\Omega \cr 
         &\deg\phi \cr 
         &\deg\pi \cr
         &\deg\bar\psi\cr
         &\deg\bar\chi \cr}
\eqalign{&\!\!=\phantom{-}2, \cr 
         &\!\!=\phantom{-}1, \cr 
         &\!\!=\phantom{-}0, \cr 
         &\!\!=-1, \cr
         &\!\!=\phantom{-}0, \cr}
\eqno(1.3)
$$
\footnote{}{}
\footnote{${}^2$}{The generators $\alpha$, $\beta$ of $W_p(V)$ are of the 
form $\alpha^it_i$, $\beta^it_i$, where $\{t_i\}$ is a basis of $V$ 
and $\{\alpha^i\}$, $\{\beta^j\}$ are the bases of $S^1(V^\vee)$, 
$A^1(V^\vee)$ dual to $\{t_i\}$. The generators $q$, $\psi$ of 
$\Omega^*(X)$ correspond to $x$, $dx$ for any local coordinate $x$ 
of $X$. See ref. \ref{9}.}. The action of the derivations $j(\xi)$, 
$l(\xi)$, $\xi\in\Lie{\cal G}$, and $s$ on the generators is given by
$$
\eqalign{&j(\xi)\omega \cr 
         &l (\xi)\omega \cr 
         &s\omega \cr}
\eqalign{&\!\!=\xi, \cr 
         &\!\!=-[\xi,\omega], \cr 
         &\!\!=\Omega-(1/2)[\omega,\omega], \cr}
\qquad\qquad
\eqalign{&j(\xi)\Omega \cr 
         &l(\xi)\Omega \cr 
         &s\Omega \cr}
\eqalign{&\!\!=0, \cr 
         &\!\!=-[\xi,\Omega], \cr 
         &\!\!=-[\omega,\Omega], \cr}
\eqno(1.4)
$$
$$
\eqalign{&j(\xi)p \cr 
         &l(\xi)p \cr 
         &s p  \cr}
\eqalign{&\!\!=0, \cr 
         &\!\!=C(p)\xi, \cr 
         &\!\!=\phi+C(p)\omega, \cr}
\qquad\qquad
\eqalign{&j(\xi)\phi \cr
         &l(\xi)\phi \cr 
         &s\phi \cr}
\eqalign{&\!\!=0, \cr 
         &\!\!=\phi\partial C(p)\xi, \cr 
         &\!\!=-C(p)\Omega-\phi\partial C(p)\omega, \cr}
\eqno(1.5)
$$
$$
\eqalign{&j(\xi)\rho \cr 
         &l(\xi)\rho \cr 
         &s\rho \cr}
\eqalign{&\!\!=0, \cr 
         &\!\!=\xi_{\cal U}{}^\vee\rho, \cr 
         &\!\!=\pi+\omega_{\cal U}{}^\vee\rho, \cr}
\qquad\qquad
\eqalign{&j(\xi)\pi \cr 
         &l(\xi)\pi \cr 
         &s\pi \cr}
\eqalign{&\!\!=0, \cr 
         &\!\!=\xi_{\cal U}{}^\vee\pi, \cr 
         &\!\!=-\Omega_{\cal U}{}^\vee\rho+\omega_{\cal U}{}^\vee\pi, \cr}
\eqno(1.6)
$$
$$
\eqalign{&j(\xi)\bar\Omega \cr 
         &l(\xi)\bar\Omega \cr 
         &s\bar\Omega \cr}
\eqalign{&\!\!=0, \cr 
         &\!\!=-[\xi,\bar\Omega], \cr 
         &\!\!=\bar\psi-[\omega,\bar\Omega], \cr}
\qquad\qquad
\eqalign{&j(\xi)\bar\psi \cr 
         &l(\xi)\bar\psi \cr 
         &s\bar\psi \cr}
\eqalign{&\!\!=0,\cr 
         &\!\!=-[\xi,\bar\psi], \cr 
         &\!\!=[\Omega,\bar\Omega]-[\omega,\bar\psi], \cr}
\eqno(1.7)
$$
$$
\eqalign{&j(\xi)\bar\lambda \cr 
         &l(\xi)\bar\lambda \cr 
         &s\bar\lambda \cr}
\eqalign{&\!\!=0, \cr 
         &\!\!=-[\xi,\bar\lambda], \cr 
         &\!\!=\bar\chi, \cr}
\qquad\qquad
\eqalign{&j(\xi)\bar\chi \cr 
         &l(\xi)\bar\chi \cr 
         &s\bar\chi \cr}
\eqalign{&\!\!=-[\xi,\bar\lambda],\cr 
         &\!\!=-[\xi,\bar\chi], \cr 
         &\!\!=0, \cr}
\eqno(1.8)
$$
in the so called intermediate or BRST model \ref{15-18,23}
\footnote{}{}
\footnote{${}^3$}{For a right action $R_X:X\times G\to X$ of a group $G$ on
a manifold $X$, we shall write interchangeably $R_X(x,g)$, $R_{Xx}(g)$,
$R_{Xg}(x)$ and $xg_X$, for $x\in X$ and $g\in G$. Similarly, we shall denote 
by $xt_X$ the tangent map $TR_{Xx}(1)t$ with $t\in{\goth g}$ and by 
$ug_X$ the tangent map $TR_{Xg}(x)u$ with $u\in T_xX$. Recall that,
for fixed $t\in{\goth g}$, the map $x\in X\to xt_X\in T_xX$ is the vertical
vector field associated with $t$.}.
Above, $[\cdot,\cdot]$ denotes the Lie bracket 
of $\L ie{\cal G}$ and $C(p)\xi$ is the vertical vector field of $\cal P$ 
associated with $\xi\in\Lie{\cal G}$. The cohomology of the complex
$({\cal W},s)$ will be called ordinary $s$ cohomology below.

Besides $(\Lie{\cal G},{\cal W},j,l,s)$, one can consider the 
operation $(\Lie{\cal G},{\cal W}_{\rm equiv}({\cal P}),j,l,s)$, where
${\cal W}_{\rm equiv}({\cal P})$ is given by the right hand side of (1.2) 
with the last three tensor factors deleted and the derivations 
$j$, $l$ and $s$ are the same as above. The basic cohomology of the complex
$({\cal W}_{\rm equiv}({\cal P}),s)$ is called equivariant $s$ cohomology 
of $\cal P$. 

To construct the topological field theory action, it is necessary to provide 
the relevant field spaces with invariant metrics as follows.
\par\noindent
\item{$i$)} An $\Ad{\cal G}$ invariant metric $(\cdot,\cdot)$ on 
$\Lie{\cal G}$ together with the induced right $\cal G$ invariant metric 
$(\cdot,\cdot)_{\cal G}$
\footnote{}{}
\footnote{${}^4$}{For a group $G$ with Lie algebra $\goth g$ equipped
with an $\Ad G$ invariant metric $(\cdot,\cdot)_{\goth g}$, the induced 
metric $(\cdot,\cdot)_G$ on $G$ is defined by $(\delta g_1,\delta g_2)_{G,g}=
(\zeta(g)\delta g_1,\zeta(g)\delta g_2)_{\goth g}$, for $g\in G$ and 
$\delta g_1,~\delta g_2\in T_gG$, where $\zeta$ is the Maurer--Cartan form of 
$G$.}.
\item{$ii$)} A $\cal G$ invariant metric $(\cdot,\cdot)_{\cal P}$ on $\cal P$.
\item{$iii$)} A $\cal G$ invariant metric $(\cdot,\cdot)_{\cal U}$ on 
$\cal U$.

The action of cohomological topological 
field theory consists of three sectors: 
the Mathai--Quillen localization sector, the Weil projection sector and the 
Faddeev--Popov gauge fixing sector:
$$
S=S_{MQ}+S_W+S_{FP}.
\eqno(1.9) 
$$
The three contributions are all $s$-exact, i. e.  obtained by applying 
$s$ to gauge fermions $\Phi_{MQ}$, $\Phi_W$, $\Phi_{FP}$ 
of degree $-1$, respectively:
$$ 
-S_{MQ}=s\Phi_{MQ},
\eqno(1.10)
$$
$$
j(\xi)\Phi_{MQ}=0, \qquad \qquad  l(\xi)\Phi_{MQ}=0,
\eqno(1.11) $$
$$
-S_W=s\Phi_W,
\eqno(1.12) 
$$
$$
j(\xi)\Phi_W=0,  \qquad \qquad l(\xi)\Phi_W=0,
\eqno(1.13)
$$
$$ 
-S_{FP}=s\Phi_{FP},
\eqno(1.14)
$$
$$
j(\xi)\Phi_{FP}\ne 0 \qquad \qquad l(\xi)\Phi_{FP}\ne 0.
\eqno(1.15) 
$$
The fact that $\Phi_{FP}$ is not annihilated by $j(\xi)$, $l(\xi)$ 
is actually required by gauge fixing.

In the standard formulation of cohomological topological field theory, 
the Mathai--Quillen gauge fermion is given by
$$
\Phi_{MQ}=-i\langle\rho,K(p)\rangle_{\cal U}
-{1\over 4}(\rho,\pi)_{\cal U}{}^\vee.
\eqno(1.16)
$$
Here, $(\cdot,\cdot)_{\cal U}{}^\vee$ is the metric on ${\cal U}^\vee$ 
canonically induced by $(\cdot,\cdot)_{\cal U}, \langle \cdot,\cdot 
\rangle_{\cal U}$ is the duality pairing of ${\cal U}^\vee$ and $\cal U$ 
and $K:{\cal P}\to{\cal U}$ is a $\cal G$ equivariant map. $K$ 
defines a section of the vector bundle 
${\cal E}={\cal P}\times_{\cal G} {\cal U}$ on 
$\cal M$. In practice, $K$ may be defined only for a subspace of 
$\cal P$ of the form $\pi_{\cal P}{}^{-1}(O)$ for some open neighborhood $O$ 
of $\cal M$. This will eventually entail Gribov type problems in the 
topological quantum field theory. From (1.5), (1.6), (1.10) and (1.16), one 
finds
$$
-S_{MQ}=-{1\over4}(\pi,\pi)_{\cal U}{}^\vee
-i\langle\pi,K(p)\rangle_{\cal U}+i\langle\rho,
\phi\partial K(p)\rangle_{\cal U}
-{1\over4}(\rho,\Omega_{\cal U}{}^\vee\rho)_{\cal U}{}^\vee.
\eqno(1.17) 
$$
As well known,
$$
I_{MQ}=\int d\rho d\pi e^{-S_{MQ}}
\eqno(1.18) 
$$
defines an equivariant $s$ cohomology class of $\cal P$ independent 
from the metric $(\cdot,\cdot)_{\cal U}$. 
Rescaling $(\cdot,\cdot)_{\cal U}$ into $u(\cdot,\cdot)_{\cal U}$, $u>0$, 
and taking the limit $u\to+\infty$, one has
$$
I_{MQ}\equiv \delta(K(p))\delta(\phi\partial K(p)),
\eqno(1.19)
$$
where $\equiv$ denotes equivalence in equivariant $s$ cohomology \ref{23}
\footnote{}{}
\footnote{${}^5$}{According to the conventions used in this paper, 
for two boson/fermion pairs of fields $l$, $\lambda$ 
and $a$, $\alpha$ valued respectively in the $n$ dimensional vector spaces 
$E$ and $F$ related by a pairing $\langle\cdot,\cdot\rangle$, the measures
$dld\lambda$ and $dad\alpha$ are such that $\int dld\lambda 
e^{i\langle l,a\rangle+i\langle\lambda,\alpha\rangle}
=\delta(a)\delta(\alpha)$ and
$\int dad\alpha \delta(a)\delta(\alpha)=(2\pi i)^{-n}(-1)^{n(n-1)\over 2}$.
If $L:E\to E$ is an invertible linear map, then $d(Ll)d(L\lambda)=
\sgn\dtr L dld\lambda$. Similarly, if $M:F\to F$ is an invertible linear map, 
then $\delta(Ma)\delta(M\alpha)=\sgn\dtr M \delta(a)\delta(\alpha)$.}. 

The Weil gauge fermion is given by
$$
\Phi_W=-i(\bar\Omega,C^\dagger(p)\phi),
\eqno(1.20)
$$
where $C^\dagger(p)$ denotes the adjoint of $C(p)$ with respect to the metrics 
$(\cdot,\cdot)$ on $\Lie{\cal G}$ and $(\cdot,\cdot)_{\cal G}$ on $\cal G$.
A straightforward calculation using standard properties of $C(p)$, (1.5), 
(1.7), (1.12) and (1.20) yields
$$
-S_W=-i(\bar \psi,C^\dagger(p)\phi)+
i(\bar\Omega,C^\dagger C(p)\Omega-\phi\partial C^\dagger(p)\phi).
\eqno(1.21) 
$$
The integral 
$$
I_W=\int d\bar\Omega d\bar\psi e^{-S_W}
\eqno(1.22)
$$
defines an equivariant $s$ cohomology class of $\cal P$ independent from
$(\cdot,\cdot)$. Indeed, $I_W$ is distributional \ref{23}
$$
I_W\equiv\delta\bigg(-{1\over C^\dagger C(p)}C^\dagger(p)\phi\bigg)
\delta\bigg(\Omega-{1\over C^\dagger C(p)}\phi\partial C^\dagger(p)\phi\bigg).
\eqno(1.23) 
$$

The Faddeev--Popov gauge fermion is given by
$$
\Phi_{FP}=-i(\bar\lambda,\Sigma(p))-{1\over4t}(\bar\lambda,\bar\chi).
\eqno(1.24) $$
Here, $t>0$ is a gauge fixing parameter.
$\Sigma:\pi_{\cal P}{}^{-1}(O)\to\Lie{\cal G}$ 
is a gauge fixing function, $O$ being some open neighborhood 
of $\cal M$. It is in general defined only locally in $\cal P$ because
of the usual Gribov problem. From (1.5), (1.8), (1.14) and (1.24), one has
$$
-S_{FP}=-{1\over4t}(\bar\chi,\bar\chi)-i(\bar\chi,\Sigma(p))
+i(\bar\lambda,(\phi+C(p)\omega)\partial\Sigma(p)).
\eqno(1.25) 
$$
The Faddeev--Popov gauge fixing factor in the functional integral is
$$
I_{FP}=\int d\bar\lambda d\bar\chi e^{-S_{FP}}.
\eqno(1.26) 
$$
It defines an ordinary $s$ cohomology class on $\cal P$ independent from 
$(\cdot,\cdot)$ and $t$. Taking the limit $t\to\infty$, one finds
$$
I_{FP} \simeq \delta(\Sigma(p))\delta((\phi+ 
C(p)\omega)\partial\Sigma(p)),
\eqno(1.27) 
$$
where $\simeq$ denotes equivalence in ordinary $s$ cohomology \ref{23}.

Let ${\cal O}={\cal O}(p,\phi,\Omega)$ be an equivariant $s$ cohomology 
class of $\cal P$. Then, the product $I_{MQ}{\cal O}$ is an equivariant $s$ 
cohomology class of $\cal P$ as well, as $I_{MQ}$ is. By the Weil homomorphism 
theorem \ref{31}, it yields a de Rham cohomology class of $\cal M$ upon 
replacing $\omega$, $\Omega$ by $\upsilon$, $\Upsilon$, where $\upsilon$ is 
a connection on $\cal P$ and $\Upsilon=s\upsilon+{1\over2}[\upsilon,\upsilon]$ 
is its curvature. Further, by Cartan's third theorem, such a class does not 
depend on the choice of $\upsilon$. Now, consider the functional integral
$$
I_{\cal O}=\int dp d\phi\int d\omega d\Omega I_{MQ}I_WI_{FP}{\cal O}.
\eqno(1.28) 
$$
It can be shown that $I_W$ can be recast as 
$$
I_W\equiv
\delta(\omega-\upsilon(p))\delta(\Omega-\Upsilon(p)),
\eqno(1.29)
$$
where $\upsilon$ is the metric connection on $\cal P$ associated with 
the metrics $(\cdot,\cdot)$ and $(\cdot,\cdot)_{\cal P}$,
$$
\upsilon(p)={1\over{C^\dagger C(p)}}C^\dagger(p)sp,
\eqno(1.30)
$$
and $\Upsilon$ is its curvature. In this way the $\omega$, $\Omega$ 
integration 
implements the Weil homomorphism. Assuming that the Gribov problems can be 
solved, one finds then that
$$
I_{\cal O}=\int_{{\cal M}}\vartheta_K\wedge\varphi_{\cal O},
\eqno(1.31) 
$$
where $\vartheta_K$ is the Poincar\'e dual of the submanifold ${\cal M}_K$ 
of $\cal M$ defined by the equivariant condition $K(p)=0$ and 
$\varphi_{\cal O}$ is a form on $\cal M$ corresponding in one--to--one 
fashion 
to $\cal O$. Of course, the above formula is only formal since $\cal M$ is 
in general infinite dimensional and non compact.

For the reader familiar with Donaldson--Witten topological
Yang-Mills theory, which we shall discuss in more detail in sect. 5,  
but less familiar with the formalism used here, it may be useful to keep 
in mind the following identifications. $\cal P$ corresponds to the space 
$\cal A$ of all gauge connections $a$. $\cal G$ is the gauge group. 
$K(p)$ is the antiself--dual part $F_-(a)$ of the 
curvature of $a$. $C(p)$ is the gauge covariant derivative 
$D(a)$ acting on Lie valued scalar fields. $C(p)^\dagger$
and $\partial K(p)$ are respectively the covariant divergence $D(a)^\dagger$
and the antiself--dual part $D_-(a)$ of the gauge covariant derivative 
acting on Lie valued vector fields.
\vskip .4cm
\par\noindent
{\bf 2. The Case of Non Free Group Actions}
\vskip .4cm
\par
The standard framework expounded in the previous section assumes 
that the action of $\cal G$ on $\cal P$ is free. In that 
case, the operator $C(p)$ has no zero modes and the standard connection 
$\upsilon$, given by (1.30), is well defined.
In field theory such an ideal situation rarely occurs and most 
group actions are not free. The following is a proposal for a general recipe 
for the treatment of this problem.

Let $\cal A$ be a manifold carrying a non free right action of a group
$\widehat{\cal G}$. For a given $a\in{\cal A}$, its $\widehat{\cal G}$ 
stability subgroup $\widehat{\cal G}(a)$ is therefore generally non trivial.
In this way, $\cal A$ may be partitioned into a disjoint union of 
$\widehat{\cal G}$ invariant subspaces ${\cal A}^x$ in one--to--one 
correspondence with the conjugacy classes $x$ of the stability 
group \ref{32}. By definition, the $\widehat{\cal G}$ orbit space of 
$\cal A$ is 
$$
{\cal B}={\cal A}/\widehat{\cal G}.
\eqno(2.1)
$$
The space $\cal B$ is not a manifold, since the $\widehat{\cal G}$ 
action is not free. It is rather a stratified space consisting of
strata ${\cal B}^x$ corresponding to the $\widehat{\cal G}$ invariant 
subspaces ${\cal A}^x$ in the quotient by the $\widehat{\cal G}$ 
action \ref{32}. Each stratum is a true manifold.

Let $\widehat{\cal Z}$ be the invariant subgroup of $\widehat{\cal G}$ acting 
trivially on $\cal A$. Then, for any $a\in{\cal A}$, $\widehat{\cal G}(a)
\supset\widehat{\cal Z}$. The elements $a\in{\cal A}$ for which
$\widehat{\cal G}(a)=\widehat{\cal Z}$ exactly are called irreducible and 
form a subspace ${\cal A}^*$ \ref{32}. The corresponding $\widehat{\cal G}$ 
orbits $n$ are called regular and span a stratum ${\cal B}^*$. 
The elements $a\in{\cal A}\setminus{\cal A}^*$ are called reducible and 
correspond to strata of singular orbits $n\in{\cal B}\setminus
{\cal B}^*$ \ref{32}.
\vskip .2cm
{\it The manifold $\cal P$ and its structure.} 
\vskip .2cm
In order to apply the framework of sect. 1, one cannot use $\cal A$. We
propose to substitute $\cal A$ by the space
$$
{\cal P}={\cal N}\times{\cal A}\times{\cal G},
\eqno(2.2) 
$$
where $\cal N$ is a stratum of $\cal B$ fixed once and for all and 
$\cal G$ is the subgroup of 
$\widehat{\cal G}$ of the elements path connected to the identity.
$\cal P$ carries a natural right $\cal G$ action. This is the trivial 
action on $\cal N$, the given right action on $\cal A$ and the right 
multiplication action on $\cal G$. Since the latter is free, the action 
on $\cal P$ is free as well. $\cal P$ is thus a principal $\cal G$ bundle. 
Its base $\cal M$ is
$$
{\cal M}={\cal P}/{\cal G}\simeq{\cal N}\times{\cal A}.
\eqno(2.3)
$$

From (2.2), it follows that, for $p=(n,a,g)\in{\cal P}$, 
$T_p({\cal P})=T_n({\cal N})\oplus T_a({\cal A})\oplus T_g({\cal G})$.
Correspondingly, the vertical vector fields of $\cal P$ are of the form 
$$
C(p)\xi=0\oplus D(a)\xi\oplus c(g)\xi,
\eqno(2.4)
$$
for $\xi\in\Lie{\cal G}$. Here, $D(a)\xi$ and 
$c(g)\xi$ are the vertical vector fields of $\cal A$ and $\cal G$, the 
latter seen as a principal $\cal G$ bundle on the singleton manifold. 
Recall that $c(g)\xi$ is also the left invariant vector field 
of $\cal G$ corresponding to $\xi$.

The graded algebra $\Omega^*({\cal P})$ is given by
$$
\Omega^*({\cal P})=\Omega^*({\cal N})\hat\otimes 
\Omega^*({\cal A})\hat\otimes\Omega^*({\cal G}).
\eqno(2.5) 
$$
So, the generators $p$ and $\phi$ of $\Omega^*({\cal P})$
have the following structure
$$
p=(n,a,g),\quad\phi=\theta\oplus\psi\oplus\epsilon,
\eqno(2.6)
$$
where the pairs $n,\theta$, $a,\psi$ and $g,\epsilon$ are the generators 
of $\Omega^*({\cal N})$, $\Omega^*({\cal A})$ and $\Omega^*({\cal G})$,
respectively. (1.5) further implies
$$
\eqalign{&j(\xi)n \cr 
         &l(\xi)n \cr 
         &s n   \cr}
\eqalign{&\!\!=0, \cr 
         &\!\!=0, \cr 
         &\!\!=\theta, \cr}
\qquad\qquad
\eqalign{&j(\xi)\theta \cr
         &l(\xi)\theta \cr 
         &s\theta \cr}
\eqalign{&\!\!=0, \cr 
         &\!\!=0, \cr 
         &\!\!=0, \cr}
\eqno(2.7)
$$
$$
\eqalign{&j(\xi)a \cr 
         &l(\xi)a \cr 
         &s a  \cr}
\eqalign{&\!\!=0, \cr 
         &\!\!=D(a)\xi, \cr 
         &\!\!=\psi+D(a)\omega, \cr}
\qquad\qquad
\eqalign{&j(\xi)\psi \cr
         &l(\xi)\psi \cr 
         &s\psi \cr}
\eqalign{&\!\!=0, \cr 
         &\!\!=\psi\partial D(a)\xi, \cr 
         &\!\!=-D(a)\Omega-\psi\partial D(a)\omega, \cr}
\eqno(2.8)
$$
$$
\eqalign{&j(\xi)g \cr 
         &l(\xi)g \cr 
         &s g   \cr}
\eqalign{&\!\!=0, \cr 
         &\!\!=c(g)\xi, \cr 
         &\!\!=\epsilon+c(g)\omega, \cr}
\qquad\qquad
\eqalign{&j(\xi)\epsilon \cr
         &l(\xi)\epsilon \cr 
         &s\epsilon\cr}
\eqalign{&\!\!=0, \cr 
         &\!\!=\epsilon\partial c(g)\xi, \cr 
         &\!\!=-c(g)\Omega-\epsilon\partial c(g)\omega, \cr}
\eqno(2.9)
$$
for $\xi\in\Lie{\cal G}$.
\vskip .2cm
{\it Localization and the structure of the vector space $\cal U$.} 
\vskip .2cm

The $\cal G$ orbit space of $\cal P$, $\cal M$, is given by (2.3). 
The relevant orbit space is however $\cal N$, the chosen stratum of 
the $\widehat{\cal G}$ orbit space of $\cal A$. 
The extra degrees of freedom contained in $\cal M$ are eliminated 
by suitably modifying localization by adding extra equivariant localization 
conditions, which localize $\cal M$ to $\cal N$. This is achieved by, 
roughly speaking, mimicking background gauge fixing in $\cal A$
as follows. An open covering 
$\{U_\alpha\}$ of $\cal N$ and a lift $a_\alpha:U_\alpha\to{\cal A}$ 
of each $U_\alpha$ to $\cal A$ are chosen. The lift must be 
such that, for $n\in U_\alpha$, $a_\alpha(n)$ belongs to the gauge orbit $n$.
An equivariant localization condition driving $a\in{\cal A}$ to $a_\alpha(n)$ 
for a fixed $n\in{\cal N}$ is then imposed. It is not obvious 
that this can be done at all but, as will be shown in due course, it can. 
The extra localization conditions being defined only locally on 
$\cal N$, a Gribov problem is involved.

To implement localization in the way indicated according to the general
method described in sect. 1, one needs therefore two localizing
functionals. The first functional is valued in some vector 
space $\cal E$ and is employed to implement ordinary localization.
The second functional is valued in another vector space 
$\cal F$ and serve the purpose of localizing $\cal M$ to $\cal N$. 
Thus, the space $\cal U$ has a direct sum decomposition 
$$
{\cal U}={\cal E}\oplus{\cal F}.
\eqno(2.10)
$$
$\cal E$ and $\cal F$ carry respectively a 
left $\widehat{\cal G}$ action and a left $\cal G$ action 
inducing the left $\cal G$ action on $\cal U$.

The vector space $\cal E$ may be quite arbitrary. $\cal F$ can be 
identified with the typical fiber of the tangent bundle $T{\cal A}$.
In fact, in order to drive $a\in{\cal A}$ to the background value 
$a_\alpha(n)$, one needs an $\cal F$ valued functional on $\cal A$ that 
vanishes and is locally invertible near $a_\alpha(n)$. This entails 
immediately the existence of linear isomorphism of $T{\cal A}$ and $\cal F$. 

The graded algebra $W_{-1}({\cal U}^\vee)$ is given by
$$
W_{-1}({\cal U}^\vee)=
W_{-1}({\cal E}^\vee)\hat\otimes W_{-1}({\cal F}^\vee)
\eqno(2.11)
$$
Correspondingly, the generators $\rho$, $\pi$ of $W_{-1}({\cal U}^\vee)$
have the structure
$$
\rho=\varrho\oplus\bar\omega,\quad\pi=\varpi\oplus\bar\tau,
\eqno(2.12)
$$
where the pairs $\varrho,\varpi$ and $\bar\omega,\bar\tau$ are the 
generators of $W_{-1}({\cal E}^\vee)$ and $W_{-1}({\cal F}^\vee)$,
respectively. From (1.6), one has
$$
\eqalign{&j(\xi)\varrho \cr 
         &l(\xi)\varrho \cr 
         &s\varrho \cr}
\eqalign{&\!\!=0, \cr 
         &\!\!=\xi_{\cal E}{}^\vee\varrho, \cr 
         &\!\!=\varpi+\omega_{\cal E}{}^\vee\varrho, \cr}
\qquad\qquad
\eqalign{&j(\xi)\varpi \cr 
         &l(\xi)\varpi \cr 
         &s\varpi \cr}
\eqalign{&\!\!=0, \cr 
         &\!\!=\xi_{\cal E}{}^\vee\varpi, \cr 
         &\!\!=-\Omega_{\cal E}{}^\vee\varrho+
\omega_{\cal E}{}^\vee\varpi, \cr}
\eqno(2.13)
$$
$$
\eqalign{&j(\xi)\bar\omega \cr 
         &l(\xi)\bar\omega \cr 
         &s\bar\omega \cr}
\eqalign{&\!\!=0, \cr 
         &\!\!=\xi_{\cal F}{}^\vee\bar\omega, \cr 
         &\!\!=\bar\tau+\omega_{\cal F}{}^\vee\bar\omega, \cr}
\qquad\qquad
\eqalign{&j(\xi)\bar\tau \cr 
         &l(\xi)\bar\tau \cr 
         &s\bar\tau \cr}
\eqalign{&\!\!=0, \cr 
         &\!\!=\xi_{\cal F}{}^\vee\bar\tau, \cr 
         &\!\!=-\Omega_{\cal F}{}^\vee\bar\omega
+\omega_{\cal F}{}^\vee\bar\tau, \cr}
\eqno(2.14)
$$
for $\xi\in\Lie{\cal G}$.
\vskip .2cm
{\it The natural gauge fixing.} 
\vskip .2cm
The $\cal G$ orbit space of $\cal P$, 
$\cal M$, is naturally realized as a gauge slice in $\cal P$ as 
$$
{\cal M}\simeq\{(n,a,g)\in{\cal P}\vert g=1\},
\eqno(2.15) 
$$
on account of (2.3).
There is therefore a natural gauge fixing condition free of Gribov problems:
$$
g=1\quad\hbox{for}\quad(n,a,g)\in{\cal P}.
\eqno(2.16) 
$$
This will be used below. 
\vskip .2cm
{\it Metric structures.} 
\vskip .2cm
By the remark above (2.4), $\cal P$ can be given a $\cal G$ invariant 
metric of the form
$$
(\delta p_1,\delta p_2)_{{\cal P},p} 
=(\delta n_1,\delta n_2)_{{\cal N},n}+(\delta a_1,\delta a_2)_{{\cal A},a}
+\mu^2(\delta g_1,\delta g_2)_{{\cal G},g},
\eqno(2.17) 
$$
for $p=(n,a,g)\in{\cal P}$ and 
$\delta p_1=\delta n_1\oplus\delta a_1\oplus\delta g_1,\delta p_2
=\delta n_2\oplus\delta a_2\oplus\delta g_2\in T_p({\cal P})$. 
Above, $(\cdot,\cdot)_{\cal N}$ is any metric on $\cal N$, whose choice will 
not matter. $(\cdot,\cdot)_{\cal A}$ is a right $\widehat{\cal G}$ 
invariant metric on $\cal A$. $(\cdot,\cdot)_{\cal G}$ is the  
right $\cal G$ invariant metric on $\cal G$ induced by 
an $\Ad\widehat{\cal G}$ invariant metric $(\cdot,\cdot)$ on 
$\Lie{\cal G}$ (see footnote 4). 
$\mu^2>0$ is a parameter and, as perhaps suggested by the notation, 
will work as an infrared cutoff.

By (2.4), the adjoint of $C(p)$ with respect to the metrics $(\cdot,\cdot)$
and $(\cdot,\cdot)_{\cal P}$ is given by
$$
C^\dagger(p)\delta p=D^\dagger(a)\delta a+\mu^2c^\dagger(g)\delta g,
\eqno(2.18) 
$$
for $\delta p=\delta n\oplus\delta a\oplus\delta g\in T_p({\cal P})$.
From (2.4) and (2.18), one has thus
$$
C^\dagger C(p)\xi=D^\dagger D(a)\xi+\mu^2\xi,
\eqno(2.19)
$$
for $\xi\in\Lie{\cal G}$. In deducing this relation, one uses
the fact that the metric $(\cdot,\cdot)_{\cal G}$ is such that 
$(c(g)\xi,c(g)\eta)_{G,g}=(\xi,\eta)$ (see footnote 4), so that 
$c^\dagger c(g)=1_{\Lie{\cal G}}$.

It should be noted that the left invariant Maurer--Cartan form 
$\zeta=g^{-1}sg$ of $\cal G$ is given by
$$
\eqalign{
\zeta&=c^\dagger(g)sg \cr
     &=c^\dagger(g)\epsilon+\omega,\cr}
\eqno(2.20) 
$$
by (2.9) and the relation $c^\dagger c(g)=1_{\Lie{\cal G}}$. Hence,
$$
\eqalign{&j(\xi)\zeta \cr
         &l(\xi)\zeta \cr 
         &s\zeta\cr}
\eqalign{&\!\!=\xi, \cr 
         &\!\!=-[\xi,\zeta], \cr 
         &\!\!=-(1/2)[\zeta,\zeta], \cr}
\eqno(2.21)
$$
from standard Lie group theory.

The spaces $\cal E$ and $\cal F$ carry respectively a
$\widehat{\cal G}$ invariant metric $(\cdot,\cdot)_{\cal E}$
and a $\cal G$ invariant $(\cdot,\cdot)_{\cal F}$. To these, there 
corresponds a $\cal G$ invariant metric on $\cal U$ in obvious 
fashion
$$
(u_1,u_2)_{\cal U}=(e_1,e_2)_{\cal E}+(f_1,f_2)_{\cal F},
\eqno(2.22)
$$
for $u_1=e_1\oplus f_1,u_2=e_2\oplus f_2\in{\cal U}$.
\vskip .4cm
\noindent
{\bf 3. The Basic Construction}
\vskip .4cm
\par
The task facing us now consists in applying the general formalism of sect. 1 
to the setting described in sect. 2.
To this end the use of the intermediate model of topological symmetry,
in which the fields, the functional measures and the $\delta$ functions 
always appear in `supersymmetric' boson/fermion pairs, is crucial
\ref{33}.
\vskip .2cm
{\it The Mathai--Quillen localization sector.} 
\vskip .2cm
Localization is achieved via a collection of maps 
$K_\alpha:O_\alpha\to{\cal U}$ associated with an open covering 
$\{O_\alpha\}$ of $\cal P$. The sets $O_\alpha$ are of the
form $O_\alpha=U_\alpha\times{\cal A}
\times{\cal G}$, where $\{U_\alpha\}$ is an open covering of $\cal N$. 
As described in sect. 2, there are two types of localization 
conditions, implying the direct sum decomposition (2.10). Correspondingly, 
one has the decomposition 
$$
K_\alpha=F\oplus H_\alpha.
\eqno(3.1)
$$
$F:{\cal P}\to{\cal E}$ is a ${\cal G}$ equivariant function 
constant on $\cal N$ and $\cal G$. In fact, we shall assume more 
restrictively that 
$$
F(a\gamma_{\cal A})=\gamma^{-1}{}_{\cal E}F(a),
\qquad\qquad\gamma\in
\widehat{\cal G}.
\eqno(3.2) 
$$
The specific form of $F$ depends on the model considered.
$H_\alpha:U_\alpha\times{\cal A}\times{\cal G}\to{\cal F}$ 
has the following properties. $H_\alpha$ is $\cal G$ equivariant,
$$
H_\alpha(n,a\gamma_{\cal A},g\gamma)
=\gamma^{-1}{}_{\cal F}H_\alpha(n,a,g),
\qquad\qquad\gamma\in{\cal G},
\eqno(3.3)
$$
and satisfies
$$
s H_\alpha(n,a,g)=\iota_a(D(a)c^\dagger(g)\epsilon-\psi)+\theta\partial_n 
H_\alpha(n,a,g)+h_\alpha(n,a,g,\psi+D(a)\omega,\epsilon+c(g)\omega),
\eqno(3.4)
$$
where
$$
h_\alpha(n,a,g,\psi+D(a)\omega,\epsilon+c(g)\omega)=0
\quad\hbox{if}\quad H_\alpha(n,a,g)=0.
\eqno(3.5) 
$$
Here, $\iota: T{\cal A}\to{\cal F}$ is a map with the following properties.
For any $a\in{\cal A}$, $\iota_a\equiv\iota|_{T_a{\cal A}}$ is a linear 
isomorphism of $T_a{\cal A}$ onto $\cal F$.
$\iota$ is equivariant, i. e.  $\iota_{a\gamma_{\cal A}}
TR_{{\cal A}\gamma}(a)=\gamma^{-1}{}_{\cal F}\iota_a$, for 
$\gamma\in\widehat{\cal G}$. $\iota$ is also orthogonal, so that 
$(\iota_a\delta a_1,\iota_a\delta a_2)_{\cal F}
=(\delta a_1,\delta a_2)_{{\cal A},a}$ for any two 
$\delta a_1,\delta a_2\in T_a{\cal A}$. Finally, $\iota$ is orientation 
preserving, i. e. $\sgn\dtr\iota_a=1$.
It is easy to check that the right hand side 
of (3.4) depends on $\psi$ and $\epsilon$ only through the combinations 
$\psi+D(a)\omega$ and $\epsilon+c(g)\omega$, as required by (2.8) and (2.9).
At $g=1$, $H_\alpha(n,a,g)$ localizes $a$ to a gauge slice $a_\alpha(n)$
defined on $U_\alpha$:
$$
H_\alpha(n,a,g)|_{g=1}=0\Rightarrow a=a_\alpha(n),
\eqno(3.6) 
$$
$$
\partial_n H_\alpha (n,a,g)|_{g=1}=\iota_{a_\alpha(n)}(\partial_na_\alpha(n)).
\eqno(3.7)
$$
The above hypotheses are technical and are motivated only by their 
consequences.

By (1.17), (2.12), (2.22), (3.1) and (3.4), the Mathai--Quillen action is
$$
\eqalignno{
-S_{MQ}&=-{1\over4}(\pi,\pi)_{\cal U}{}^\vee
-i\langle\pi,K_\alpha(p)\rangle_{\cal U}
+i\langle\rho,\phi\partial K_\alpha(p)\rangle_{\cal U}- 
{1\over4}(\rho,\Omega_{\cal U}{}^\vee \rho)_{\cal U}{}^\vee &\cr
&=-{1\over4}(\varpi,\varpi)_{\cal E}{}^\vee 
-i\langle \varpi,F(a)\rangle_{\cal E}
+i\langle\varrho,\psi\partial F(a)\rangle_{\cal E}
-{1\over4}(\varrho,\Omega_{\cal E}{}^\vee\varrho)_{\cal E}{}^\vee &\cr
&\phantom{=} 
-{1\over4}(\bar\tau,\bar\tau)_{\cal F}{}^\vee
-i\langle \bar\tau, H_\alpha(n,a,g)\rangle_{\cal F}
+i\langle\bar\omega,\iota_a(D(a)c^\dagger(g)\epsilon-\psi) &\cr
&\phantom{=}+\theta\partial_nH_\alpha(n,a,g)
+h_\alpha(n,a,g,\psi,\epsilon)\rangle_{\cal F} 
-{1\over4}(\bar\omega,\Omega_{\cal F}{}^\vee\bar\omega)_{\cal F}{}^\vee. 
&(3.8)\cr} 
$$
The functional integral $\int d\rho d\pi e^{-S_{MQ}}$
defines an equivariant $s$ cohomology class independent from
$(\cdot, \cdot)_{\cal E}$ and $(\cdot, \cdot)_{\cal F}$.
By rescaling $(\cdot, \cdot)_{\cal F}$ into 
$u(\cdot,\cdot)_{\cal F}$ with $u>0$ and taking the limit $u\to+\infty$
in the integral, we obtain a representative of such class. The result is 
$$\eqalign{
\int d\rho d\pi e^{-S_{MQ}}
&\equiv\int d\varrho d\varpi 
\exp\bigg\{-{1\over4}(\varpi,\varpi)_{\cal E}{}^\vee 
-i\langle\varpi,F(a)\rangle_{\cal E}
+i\langle\varrho,\psi\partial F(a)\rangle_{\cal E}\cr
&\phantom{\equiv} 
-{1\over4}(\varrho,\Omega_{\cal E}{}^\vee\varrho)_{\cal E}{}^\vee
\bigg\}\delta(H_\alpha^*(n,a,g))
\delta(-\psi+D(a)c^\dagger(g)\epsilon+\theta\partial_n H_\alpha^*(n,a,g)), 
\cr}
\eqno(3.9) 
$$
where $\equiv$ denotes equivalence in equivariant $s$ cohomology and one has 
set $H_\alpha^*(n,a,g)=\iota_a{}^{-1}H_\alpha(n,a,g)$. Above, the formal 
rules stated in footnote 5 have been used. 
\vskip .2 cm
{\it The Weil projection sector}. 
\vskip .2cm
The action of the Weil sector is, according to (1.21), (2.6), (2.18) 
and (2.19),
$$
\eqalign{-S_W&=-i(\bar\psi,C^\dagger(p)\phi)
+i(\bar\Omega,C^\dagger C(p)\Omega
-\phi\partial C^\dagger(p)\phi)\vphantom{1\over2}\cr
&=-i(\bar\psi,D^\dagger(a)\psi+\mu^2 c^\dagger(g)\epsilon)
+i(\bar\Omega,(D^\dagger D(a)+\mu^2)\Omega
-\psi\partial D^\dagger(a)\psi 
-\mu^2\epsilon\partial c^\dagger(g)\epsilon).\vphantom{1\over2}\cr}
\eqno(3.10) 
$$
Hence, the functional integral $\int d\bar\psi d\bar\Omega e^{-S_W}$
is given by
$$
\int d\bar\Omega d\bar\psi e^{-S_W} 
\equiv\delta((D^\dagger D(a)+\mu^2)\Omega
-\psi\partial D^\dagger(a)\psi-\mu^2\epsilon\partial c^\dagger(g)\epsilon)
\delta(-(D^\dagger(a)\psi+\mu^2c^\dagger(g)\epsilon)).
\eqno(3.11)
$$
This expression defines an equivariant $s$ cohomology 
class independent from $(\cdot,\cdot)$ and $\mu ^2$. Then, its limit as 
$\mu^2\to0$, if it exists, provides a representative of such a class.
Let $q(a)$ be the orthogonal projector of $\Lie{\cal G}$ onto $\ker D(a)$.
Since $D(a)q(a)=0$ and $q(a)=q(a)^\dagger$, one has $q(a)D^\dagger(a)=0$. 
Considering $D^\dagger(a)$ as a $\Lie{\cal G}$ valued one form, one has thus  
$d_aq(a)\wedge D^\dagger(a)+q(a)\wedge d_aD^\dagger(a)=0$. Using this 
relation, one can show that $\int d\bar\psi d\bar\Omega e^{-S_W}$ 
has a formal limit as $\mu^2\to0$ and compute it. In this way, 
one finds that  
$$
\eqalignno{\int d\bar\Omega d\bar\psi e^{-S_W} 
&\equiv\delta(q(a)(\Omega-\psi\partial q(a)c^\dagger(g)\epsilon
-\epsilon\partial c^\dagger(g)\epsilon))\delta(-q(a)c^\dagger(g)\epsilon) &\cr
&\phantom{\equiv}\vphantom{\int}\times\delta((1-q(a))(D^\dagger D(a)\Omega
-\psi\partial D^\dagger(a)\psi))\delta(-(1-q(a))D^\dagger(a)\psi)~~~~~~
&(3.12)\cr}
$$
in equivariant $s$ cohomology.
\vskip .2cm
{\it The Faddeev--Popov gauge fixing sector}. 
\vskip .2cm
As indicated in sect. 2, a suitable Gribov free gauge fixing condition is 
given by (2.16). We assume 
that the gauge fixing function $\Sigma:{\cal P}\to\Lie{\cal G}$ is of the 
form
$$
\Sigma(p)=C^\dagger(p)\Xi(p)
\eqno(3.13)
$$
where $\Xi:{\cal P}\to T{\cal P}$ is a section of $T{\cal P}$ constant on 
$\cal N$ with the following structure:
$$
\Xi(a,g)=0\oplus W(a,g)\oplus c(g)\ln g.
\eqno(3.14)
$$
Here, $W:{\cal P}\to p_{\cal A}{}^*(T{\cal A})$ is a section of the 
pull--back of $T{\cal A}$ via the natural projection 
$p_{\cal A}:{\cal P}\to{\cal A}$ constant on the factor $\cal N$ of $\cal P$
and such that 
$$
W(a,g)|_{g=1}=0,
\eqno(3.15) 
$$
and 
$$
sW(a,g)=D(a)(c^\dagger(g)\epsilon+\omega)
+w(a,g)((\psi+D(a)\omega)\oplus(\epsilon+c(g)\omega)),
\eqno(3.16)
$$ 
where 
$$
w(a,g)|_{g=1}=0.
\eqno(3.17)
$$
$\ln:{\cal G}\to \Lie{\cal G}$ is the inverse of the exponential map 
$\exp:\Lie{\cal G}\to{\cal G}$. It is defined only near $g=1$ so that $K$ is 
not defined far away from $g=1$. The solution of this problem will be 
provided later in this section when discussing renormalizability. 
Using (1.25), (2.18)--(2.19), (3.13), (3.14), (3.16) and the relations
$c^\dagger c(g)=1_{\Lie{\cal G}}$ and 
$s(c(g)\ln g)=c(g)(c^\dagger(g)\epsilon+\omega)
+o(g)(\epsilon+c(g)\omega)$, where $o(g)|_{g=1}=0$, one has
$$
\eqalignno{-S_{FP}&=-{1\over4t}(\bar\chi,\bar\chi)-i(\bar\chi,
\Sigma(p))+i(\bar\lambda,(\phi+C(p)\omega)\partial\Sigma(p)) &\cr
&=-{1\over4t}(\bar\chi,\bar\chi)-i(\bar\chi,D^\dagger (a)W(a,g)+\mu^2\ln g)
+i(\bar\lambda,(D^\dagger D(a)+\mu^2)(c^\dagger(g)\epsilon+\omega) &\cr
&\phantom{=}\vphantom{1\over4t}
+D^\dagger(a)w(a,g)((\psi+D(a)\omega)\oplus(\epsilon+c(g)\omega))
+(\psi+D(a)\omega)\partial D^\dagger(a)W(a,g) &\cr
&\phantom{=}
\vphantom{1\over4t}+\mu^2(c^\dagger(g)o(g)(\epsilon+c(g)\omega)+
(\epsilon+c(g)\omega)\partial c^\dagger(g)c(g)\ln g)).&(3.18)\cr} 
$$
On general grounds, the integral $\int d\bar\lambda d\bar\chi e^{-S_{FP}}$ 
defines an $s$ cohomology class independent from $(\cdot,\cdot)$, $t$ and 
$\mu^2$. One can obtain a representative of such a class by  
taking the limit $t\to+\infty$ (Landau gauge). The limit concentrates the 
integration on the zero set of the gauge fixing function $\Sigma(p)$ 
and, since $\Sigma(a,g)=0$ implies $g=1$, it makes several terms in
expression (3.18) vanish, by (3.15), (3.17) and the relations 
$c(g)\ln g|_{g=1}=0$ and $o(g)|_{g=1}=0$. One has then
$$
\int d\bar\lambda d\bar\chi e^{-S_{FP}}\simeq
\delta((D^\dagger D(a)+\mu^2)(c^\dagger(g)\epsilon+\omega))
\delta(D^\dagger(a)W(a,g)+\mu^2\ln g),
\eqno(3.19)
$$
where $\simeq$ denotes equivalence in ordinary $s$ cohomology.
As an $s$ cohomology class, this expression is still independent from $\mu^2$. 
One can thus obtain a representative of the class by taking its limit as 
$\mu^2\to 0$, provided it exists. Proceeding in this way, one finds
$$
\eqalignno{
\int d\bar\lambda d\bar\chi e^{-S_{FP}}&\simeq
\delta(q(a)(c^\dagger(g)\epsilon+\omega))\delta(q(a)\ln g) &\cr
&\phantom{\simeq}\times
\delta((1-q(a))D^\dagger D(a)(c^\dagger(g)\epsilon+\omega))
\delta((1-q(a))D^\dagger(a)W(a,g)),~~~
&(3.20)\cr}
$$
where $q(a)$ is defined above. 
\vskip .2cm
{\it Putting the three sectors together}. 
\vskip .2cm
From (3.9), (3.12) and (3.20), after some simple rearrangements, 
one finds that 
$$
\eqalignno{
&\int d\rho d\pi\int d\bar\Omega d\bar\psi\int
d\bar\lambda d\bar\chi e^{-S_{MQ}-S_W-S_{FP}}&\cr
&\simeq
\int d\varrho d\varpi
\exp\bigg\{-{1\over4}(\varpi,\varpi)_{\cal E}{}^\vee
-i\langle\varpi,F(a)\rangle_{\cal E}
+i\langle \varrho,\psi\partial F(a)\rangle_{\cal E}
-{1\over4}(\varrho,\Omega_{\cal E}{}^\vee \varrho)_{\cal E}{}^\vee\bigg\} &\cr
&\phantom{\simeq}\vphantom{\bigg\{}\times
\delta(H_\alpha^*(n,a,g))
\delta(-\psi+D(a)c^\dagger(g)\epsilon+\theta\partial_n H_\alpha^*(n,a,g)) &\cr
&\phantom{\simeq}\vphantom{\bigg\{}\times
\delta(q(a)(\Omega-(1/2)[\omega,\omega]
+\theta\partial_n H_\alpha^*(n,a,g)\partial q(a)\omega))
\delta(q(a)\omega)&\cr
&\phantom{\simeq}\vphantom{\bigg\{}\times
\delta((1-q(a))(D^\dagger D(a)\Omega
-\psi\partial D^\dagger(a)\psi))\delta(-(1-q(a))D^\dagger(a)\psi) &\cr
&\phantom{\simeq}\vphantom{\bigg\{}\times
\delta(q(a)(f(g)(c^\dagger(g)\epsilon+\omega)
+\theta\partial_n H_\alpha^*(n,a,g)\partial q(a)\ln g))
\delta(q(a)\ln g) &\cr
&\phantom{\simeq}\vphantom{\bigg\{}\times
\delta((1-q(a))D^\dagger D(a)(c^\dagger(g)\epsilon+\omega))
\delta((1-q(a))D^\dagger(a)W(a,g)) &(3.21) \cr}
$$
in $s$ cohomology. Here, $f(g)$ is defined by the relation 
$f(g)(c^\dagger(g)\epsilon+\omega)=s\ln g$ and is explicitly given by 
$f(g)=\ad\ln g/(1-\exp(-\ad\ln g))$. To show the above relation,
one uses the fact that the $\delta$ functions of the Faddeev--Popov 
sector enforce the constraints $g=1$ and $c^\dagger(g)\epsilon+\omega=0$ 
and the identities $(c(g)\omega)\partial c^\dagger(g)(c(g)\omega)
=-(1/2)[\omega,\omega]$ and $(D(a)\omega)\partial q(a)\omega
=-[\omega,q(a)\omega]+q(a)[\omega,\omega]$. These follow respectively 
from the 
Maurer--Cartan equation $d_gc^\dagger(g)=-(1/2)[c^\dagger(g),c^\dagger(g)]$
and the relation $l(\xi)q(a)=-[\ad\xi,q(a)]$, $\xi\in\Lie{\cal G}$.

The $\delta$ functions appearing in the right hand side of (3.21) have
functional integral representations. The Mathai--Quillen factor is given by
$$
\eqalignno{\phantom{=}\vphantom{\bigg\{}
&\delta(H^*_\alpha(n,a,g))\delta(-\psi+D(a)c^\dagger(g)\epsilon
+\theta\partial_n H_\alpha^*(n,a,g)) &\cr
=&\lim_{b\to+\infty}\int d\bar\omega d\bar\tau
\exp\bigg\{-{1\over4b}(\bar\tau,\bar\tau)_{\cal F}{}^\vee
-i\langle \bar\tau,H_\alpha(n,a,g)\rangle_{\cal F}
+i\langle\bar\omega,\iota_a(D(a)c^\dagger(g)\epsilon-\psi) &\cr
\phantom{=}&
+\theta\partial_nH_\alpha(n,a,g)
+h_\alpha(n,a,g,\psi,\epsilon)\rangle_{\cal F} 
-{1\over4b}(\bar\omega,\Omega_{\cal F}{}^\vee\bar\omega)_{\cal F}{}^\vee
\bigg\}.&(3.22) \cr}
$$
The Weil factor is
$$
\eqalignno{\phantom{=}
&\vphantom{\bigg\{}
\delta(q(a)(\Omega-(1/2)[\omega,\omega]
+\theta\partial_n H_\alpha^*(n,a,g)\partial q(a)\omega))
\delta(q(a)\omega)
\delta((1-q(a))(D^\dagger D(a)\Omega &\cr
&\vphantom{\bigg\{}
-\psi\partial D^\dagger(a)\psi))
\delta(-(1-q(a))D^\dagger(a)\psi) &\cr
=&\int d\bar\Omega d\bar\psi 
\exp\bigg\{-i(\bar\psi,D^\dagger(a)\psi)+i(\bar\Omega,D^\dagger D(a)\Omega
-\psi\partial D^\dagger(a)\psi)\bigg\} &\cr
\phantom{=}&\vphantom{\bigg\{}
\times(2\pi i)^{d(a)}(-1)^{d(a)(d(a)-1)\over2}
\delta(q(a)\bar\Omega)\delta(q(a)(\bar\psi-[\omega,\bar\Omega]
+\theta\partial_n H_\alpha^*(n,a,g)\partial q(a)\bar\Omega)) &\cr
\phantom{=}&\vphantom{\bigg\{}
\times\delta(q(a)\omega)\delta(q(a)(\Omega-(1/2)[\omega,\omega]
+\theta\partial_n H_\alpha^*(n,a,g)\partial q(a)\omega)), &(3.23) \cr}
$$
where $d(a)=\dim\ker D(a)$. 
Finally, the Faddeev--Popov factor has the representation
$$
\eqalignno{
&\vphantom{\bigg\{}\delta(q(a)(f(g)(c^\dagger(g)\epsilon+\omega)
+\theta\partial_n H_\alpha^*(n,a,g)\partial q(a)\ln g))
\delta(q(a)\ln g) &\cr
\phantom{=}&\vphantom{\bigg\{}\times
\delta((1-q(a))D^\dagger D(a)(c^\dagger(g)\epsilon+\omega))
\delta((1-q(a))D^\dagger(a)W(a,g)) &\cr
=&\lim_{t\to+\infty}\int d\bar\lambda d\bar\chi\exp\bigg\{
-{1\over4t}(\bar\chi,\bar\chi)
-i(\bar\chi,D^\dagger(a)W(a,g))
+i(\bar\lambda,D^\dagger D(a)(c^\dagger(g)\epsilon+\omega) &\cr
\phantom{=}&
+D^\dagger(a)w(a,g)((\psi+D(a)\omega)\oplus(\epsilon+c(g)\omega))
+(\psi+D(a)\omega)\partial D^\dagger(a)W(a,g))\bigg\} &\cr
\phantom{=}&\vphantom{\bigg\{}
\times(2\pi i)^{d(a)}(-1)^{d(a)(d(a)-1)\over2}
\delta(q(a)\bar\lambda)
\delta(q(a)(\bar\chi
+\theta\partial_n H_\alpha^*(n,a,g)\partial q(a)\bar\lambda)) &\cr
\phantom{=}&\vphantom{\bigg\{}
\delta(q(a)\ln g)
\delta(q(a)(f(g)(c^\dagger(g)\epsilon+\omega)
+\theta\partial_n H_\alpha^*(n,a,g)\partial q(a)\ln g)). &(3.24)\cr}
$$
In deducing the above relations, one uses the standard properties of the 
boson/fermion $\delta$ functions, the orthogonal factorization of the 
integration measures, the relation $q(a)D^\dagger(a)$ $=$ $0$ 
and (3.15) and (3.17). 
The signs are determined according the conventions stated in footnote 5 
above. 

Next, eqs. (3.22)--(3.24) are to be substituted into eq. (3.21).
Now, the $\delta$ functions $\delta(q(a)\ln g)
\delta((1-q(a))D^\dagger(a)W(a,g))$ arising in the Faddeev--Popov sector 
enforce the identity $g=1$, by (3.13) and (3.14). The $\delta$ function 
$\delta(H_\alpha^*(n,a,1))$ resulting in this way in the Mathai-Quillen 
sector allows one to set $a=a_\alpha(n)$ everywhere else, on account 
of (3.6). 
The finite dimensional $\delta$ functions so yielded in the integral
representations (3.23) and (3.24) define $s$ cohomology classes having 
in turn the following integral representations. Set $K_\alpha(n)
=\ker D(a_\alpha(n))$ for $n\in U_\alpha$. Consider the auxiliary graded 
tensor algebra
$$
{\cal Z}_\alpha(n)=W_{-2}(K_\alpha(n))\hat\otimes W_1(K_\alpha(n))
\hat\otimes W_{-1}(K_\alpha(n))\hat\otimes W_0(K_\alpha(n)).
\eqno(3.25)
$$
The generators of the factors in the given order are
$$
\eqalign{&\bar y_\alpha(n),~\bar x_\alpha(n) \cr 
         &x_\alpha(n),~y_\alpha(n) \cr 
         &u_\alpha(n),~v_\alpha(n) \cr 
         &r_\alpha(n),~z_\alpha(n) \cr}
\qquad\qquad
\eqalign{&\deg\bar y_\alpha(n)\cr 
         &\deg x_\alpha(n)  \cr 
         &\deg u_\alpha(n)\cr
         &\deg r_\alpha(n)\cr}
\eqalign{&\!\!=-2,~ \cr 
         &\!\!=\phantom{-}1,~ \cr 
         &\!\!=-1,~ \cr
         &\!\!=\phantom{-}0,~ \cr}
\eqalign{&\deg\bar x_\alpha(n)\cr 
         &\deg y_\alpha(n)\cr 
         &\deg v_\alpha(n)\cr
         &\deg z_\alpha(n)\cr}
\eqalign{&\!\!=-1, \cr 
         &\!\!=\phantom{-}2, \cr 
         &\!\!=\phantom{-}0, \cr 
         &\!\!=\phantom{-}1. \cr}
\eqno(3.26)
$$
The derivations $j(\xi)$, $l(\xi)$, $\xi\in\Lie{\cal G}$, and $s$
are extended as follows:
$$
\eqalign{&j(\xi)f_\alpha(n) \cr 
         &l(\xi)f_\alpha(n) \cr 
         &s f_\alpha(n)   \cr}
\eqalign{&\!\!=0, \cr 
         &\!\!=0, \cr 
         &\!\!=g_\alpha(n)+\theta\partial_nf_\alpha(n), \cr}
\qquad\qquad
\eqalign{&j(\xi)g_\alpha(n) \cr
         &l(\xi)g_\alpha(n) \cr 
         &s g_\alpha(n) \cr}
\eqalign{&\!\!=0, \cr 
         &\!\!=0, \cr 
         &\!\!=\theta\partial_ng_\alpha(n). \cr}
\eqno(3.27)
$$
with $(f,g)=(\bar y,\bar x),(x,y),(u,v),(r,z)$. 
One considers next the gauge fermions
$$
\eqalignno{
\Psi_{\rm equiv}&=i(\bar y_\alpha(n),\omega)
&(3.28)\cr
\Psi_{\rm pr\hphantom{uiv}}&=-i(x_\alpha(n),\bar\Omega)
&(3.29)\cr 
\Psi_{\rm gau\hphantom{iv}}&=-i(u_\alpha(n),\ln g)
&(3.30)\cr 
\Psi_{\rm fix\hphantom{uv}}&=i(r_\alpha(n),\bar\lambda)
&(3.31)\cr} 
$$
and defines $s$ cohomology classes 
$\Delta_{\rm equiv}$, $\Delta_{\rm pr}$, $\Delta_{\rm gau}$ and 
$\Delta_{\rm fix}$ by:
$$
\eqalignno{
-S_{\rm equiv}&=s\Psi_{\rm equiv} &\cr
&=i(\bar x_\alpha(n)+\theta\partial_n\bar y_\alpha(n),\omega)
+i(\bar y_\alpha(n),\Omega-(1/2)[\omega,\omega])),\vphantom{\int}
&(3.32)
\cr
\Delta_{\rm equiv}&=\int d\bar y_\alpha(n)d\bar x_\alpha(n)e^{-S_{\rm equiv}}
&\cr
&=\delta(q(a_\alpha(n))\omega)
\delta(q(a_\alpha(n))(\Omega-(1/2)[\omega,\omega]
+\theta\partial_n(q(a_\alpha(n)))\omega))\vphantom{\int}, &(3.33) \cr
-S_{\rm pr}&=s\Psi_{\rm pr} &\cr
&=-i(y_\alpha(n)+\theta\partial_nx_\alpha(n),\bar\Omega )
+i(x_\alpha(n),\bar\psi-[\omega,\bar\Omega]),\vphantom{\int}
&(3.34) \cr
\Delta_{\rm pr}&=\int dx_\alpha(n)dy_\alpha(n)e^{-S_{\rm pr}} &\cr
&=\delta(q(a_\alpha(n))\bar\Omega)
\delta(q(a_\alpha(n))(\bar\psi-[\omega,\bar\Omega]
+\theta\partial_n(q(a_\alpha(n)))\bar\Omega))
\vphantom{\int}, &(3.35) \cr
-S_{\rm gau}&=s\Psi_{\rm gau} &\cr
&=-i(v_\alpha(n)+\theta\partial_nu_\alpha(n),\ln g)
+i(u_\alpha(n),f(g)(c^\dagger(g)\epsilon+\omega)),\vphantom{\int}
&(3.36) \cr
\Delta_{\rm gau}&=\int du_\alpha(n)dv_\alpha(n)e^{-S_{\rm gau}} &\cr
&=\delta(q(a_\alpha(n))\ln g)
\delta(q(a_\alpha(n))(f(g)(c^\dagger(g)\epsilon+\omega)
+\theta\partial_n(q(a_\alpha(n)))\ln g))\vphantom{\int},~~~~
&(3.37) \cr
-S_{\rm fix}&=s\Psi_{\rm fix} &\cr
&=i(z_\alpha(n)+\theta\partial_nr_\alpha(n),\bar\lambda)
+i(r_\alpha(n),\bar\chi),\vphantom{\int}
&(3.38) \cr
\Delta_{\rm fix}&=\int dr_\alpha(n) dz_\alpha(n)e^{-S_{\rm fix}} &\cr
&=\delta(q(a_\alpha(n))\bar\lambda)
\delta(q(a_\alpha(n))(\bar\chi
+\theta\partial_n(q(a_\alpha(n)))\bar\lambda))
\vphantom{\int}. &(3.39) \cr}
$$
where $f(g)$ is defined below eq. (3.21).

From the above discussion and from (3.33), (3.35), (3.37) and (3.39), it 
follows that the product of the finite dimensional $\delta$ functions 
appearing in (3.23) and (3.24) with $a=a_\alpha(n)$ may be substituted by the 
product $\Delta_{\rm equiv}\Delta_{\rm pr}\Delta_{\rm gau}\Delta_{\rm fix}$.
Proceeding in this way, one finds the formula 
$$
\eqalignno{
&\int d\rho d\pi\int d\bar\Omega d\bar\psi\int
d\bar\lambda d\bar\chi e^{-S_{MQ}-S_W-S_{FP}} &\cr
&\simeq\lim_{t,b\to+\infty}\int d\bar y_\alpha(n)d\bar x_\alpha(n)
\int dx_\alpha(n)dy_\alpha(n)\int du_\alpha(n)dv_\alpha(n)
\int dr_\alpha(n)dz_\alpha(n) &\cr
&\phantom{\simeq}\times(2\pi i)^{2d_{\cal N}}
\int d\varrho d\varpi\int d\bar\omega d\bar\tau
\int d\bar\Omega d\bar\psi\int d\bar\lambda d\bar\chi
e^{-S_{\rm top}}, &(3.40)\cr}
$$ 
where $d_{\cal N}=d(a_\alpha(n))$ with $n\in{\cal N}$ is a non negative 
integer constant characterizing $\cal N$ and
$$
\eqalignno{
\vphantom{\bigg\{}-S_{\rm top}&=
-{1\over4}(\varpi,\varpi)_{\cal E}{}^\vee
-i\langle\varpi,F(a)\rangle_{\cal E}
+i\langle \varrho,\psi\partial F(a)\rangle_{\cal E}
-{1\over4}(\varrho,\Omega_{\cal E}{}^\vee \varrho)_{\cal E}{}^\vee
-{1\over4b}(\bar\tau,\bar\tau)_{\cal F}{}^\vee &\cr
&\phantom{\simeq}\vphantom{\bigg\{}
-i\langle \bar\tau, H_\alpha(n,a,g)\rangle_{\cal F}
+i\langle\bar\omega,\iota_a(D(a)\zeta-\psi-D(a)\omega)
+\theta\partial_nH_\alpha(n,a,g) &\cr
&\phantom{\simeq}\vphantom{\bigg\{}
+h_\alpha(n,a,g,\psi,c(g)(\zeta-\omega))\rangle_{\cal F}
-{1\over4b}(\bar\omega,\Omega_{\cal F}{}^\vee\bar\omega)_{\cal F}{}^\vee
-i(\bar\psi,D^\dagger(a)\psi) &\cr
&\phantom{\simeq}\vphantom{\bigg\{} 
+i(\bar\Omega,D^\dagger D(a)\Omega 
-\psi\partial D^\dagger(a)\psi)
-{1\over4t}(\bar\chi,\bar\chi)
-i(\bar\chi,D^\dagger(a)W(a,g))
+i(\bar\lambda,D^\dagger D(a)\zeta  &\cr
&\phantom{\simeq}\vphantom{\bigg\{}
+D^\dagger(a)w(a,g)((\psi+D(a)\omega)\oplus c(g)\zeta)
+(\psi+D(a)\omega)\partial D^\dagger(a)W(a,g)) &\cr
&\phantom{\simeq}\vphantom{\bigg\{}
+i(\bar x_\alpha(n)+\theta\partial_n\bar y_\alpha(n),\omega)
+i(\bar y_\alpha(n),\Omega-(1/2)[\omega,\omega])
-i(y_\alpha(n)+\theta\partial_nx_\alpha(n),\bar\Omega) &\cr
&\phantom{\simeq}\vphantom{\bigg\{}
+i(x_\alpha(n),\bar\psi-[\omega,\bar\Omega])
-i(v_\alpha(n)+\theta\partial_nu_\alpha(n),\ln g)
+i(u_\alpha(n),f(g)\zeta) &\cr
&\phantom{\simeq}\vphantom{\bigg\{}
+i(z_\alpha(n)+\theta\partial_nr_\alpha(n),\bar\lambda)
+i(r_\alpha(n),\bar\chi), &(3.41)\cr}
$$
$\zeta$ being the Maurer--Cartan form given by (2.20). By 
construction, the action (3.41) is $s$ exact. Hence, in $s$ cohomology, 
the right hand side of (3.40) is independent from $t$, $b$ and the limit 
may be dropped. Note, however, that the right hand sides
of (3.21) and (3.40) are exactly equal and not merely equivalent in $s$ 
cohomology. 
$S_{\rm top}$ is the effective topological action produced by the present 
method. It must be stressed that it contains the customary topological 
action as a subsector. 


\vskip .2cm
{\it On--shell analysis}. 
\vskip .2cm
Relation (3.21) can be cast in a more transparent form as follows:
$$
\eqalignno{
&\int d\rho d\pi\int d\bar\Omega d\bar\psi\int
d\bar\lambda d\bar\chi e^{-S_{MQ}-S_W-S_{FP}} &\cr
&\simeq
\int d\varrho d\varpi
\exp\bigg\{-{1\over4}(\varpi,\varpi)_{\cal E}{}^\vee
-i\langle\varpi,F(a)\rangle_{\cal E}
+i\langle \varrho,\psi\partial F(a)\rangle_{\cal E}
-{1\over4}(\varrho,\Omega_{\cal E}{}^\vee \varrho)_{\cal E}{}^\vee\bigg\} &\cr
&\phantom{\simeq}\vphantom{\bigg\{}\times
\delta(H_\alpha^*(n,a,g))
\delta(-sa+\theta\partial_n H_\alpha^*(n,a,g))
\delta(\omega-G_{D^\dagger D}(a)D^\dagger(a)sa) &\cr
&\phantom{\simeq}\vphantom{\bigg\{}\times
\delta(s\omega-s(G_{D^\dagger D}(a)D^\dagger(a)sa))
\delta(\ln g)\delta(s\ln g), &(3.42) \cr}
$$
where $sa$ and $s\omega$ are given by (2.8) and (1.4), respectively, 
$s\ln g=f(g)\zeta$ and $G_{D^\dagger D}(a)$ is the Green function of 
$D^\dagger D(a)$ uniquely defined by the relations
$$
\eqalign{
D^\dagger D(a)G_{D^\dagger D}(a)=G_{D^\dagger D}(a)D^\dagger D(a)&=1-q(a),\cr
q(a)G_{D^\dagger D}(a)=G_{D^\dagger D}(a)q(a)&=0.\cr}
\eqno(3.43)
$$
Eq. (3.42) is obtained using the relations listed below eq. (3.21), (3.43) 
and the identities $(D(a)\omega)\partial D(a)\omega=(1/2)D(a)[\omega,\omega]$ 
and $(sa-D(a)\omega)\partial D^\dagger(a)(sa-D(a)\omega)=
(sa)\partial D^\dagger(a)sa-(sa)\partial(D^\dagger D)(a)\omega
-[\omega,D^\dagger D(a)\omega-D^\dagger(a)sa]
+(1/2)D^\dagger D(a)[\omega,\omega]$ and (3.43). These 
follow from the relations $l(\xi)D^\dagger(a)$ $=-[\xi,D^\dagger(a)]$ and 
$D^\dagger(a)[\xi,\eta]=[D^\dagger(a)\xi,D^\dagger(a)\eta]_{\cal A}$
for $\xi,\eta\in\Lie{\cal G}$, $[\cdot,\cdot]_{\cal A}$ being the 
Lie bracket on vector fields on $\cal A$. 
From (3.42), using (3.6) and (3.7), one deduces immediately that
$$
\eqalignno{
&\int d\rho d\pi\int d\bar\Omega d\bar\psi\int
d\bar\lambda d\bar\chi e^{-S_{MQ}-S_W-S_{FP}} &\cr
&\simeq
\int d\varrho d\varpi
\exp\bigg\{-{1\over4}(\varpi,\varpi)_{\cal E}{}^\vee
-i\langle\varpi,F(a)\rangle_{\cal E}
+i\langle \varrho,\psi\partial F(a)\rangle_{\cal E}
-{1\over4}(\varrho,\Omega_{\cal E}{}^\vee \varrho)_{\cal E}{}^\vee\bigg\} &\cr
&\phantom{\simeq}\vphantom{\bigg\{}\times
\delta(H_\alpha^*(n,a,1))\delta(-\psi|_{a=a_\alpha(n)}
+(1-D(a_\alpha(n))G_{D^\dagger D}(a_\alpha(n))D^\dagger(a_\alpha(n)))
\theta\partial_na_\alpha(n)) &\cr
&\phantom{\simeq}\vphantom{\bigg\{}\times
\delta(\omega-\upsilon_\alpha(n))
\delta(\Omega-\Upsilon_\alpha(n))
\delta(\ln g)\delta(f(g)\zeta), &(3.44) \cr}
$$
where 
$$
\upsilon_\alpha(n)=G_{D^\dagger D}(a_\alpha(n))D^\dagger(a_\alpha(n))
\theta\partial_n a_\alpha(n),
\eqno(3.45)
$$
$$
\Upsilon_\alpha(n)=\theta\partial_n\upsilon_\alpha(n)
+(1/2)[\upsilon_\alpha(n),\upsilon_\alpha(n)].
\eqno(3.46)
$$
Eq. (3.44) shows clearly the on--shell structure of the functional 
integral and is the starting point for the analysis of the 
Gribov problem below.
\vskip .2cm
{\it The functional integral of the topological field theory}. 
\vskip .2cm
According the general framework described in sect. 1, in a cohomological 
topological 
quantum field theory, one must integrate all the fields with action
$S_{MQ}+S_W+S_{FP}$ and insertions of an equivariant $s$ cohomology class 
$\cal O$ of $\cal P$. In the present case, one has that, in $s$ cohomology,
the $\rho,~\pi$, $\bar\Omega,~\bar\psi$, $\bar\lambda,\bar\chi$ integral
is given by (3.40)--(3.41). The resulting functional of $p,~\phi$ and 
$\omega,~\Omega$ should be further integrated with respect to those fields. 

The $n,~\theta$ integration involved by the $p,~\phi$ integration is 
problematic for two main reasons: the non compactness and infinite 
dimensionality of $\cal N$ and the lack of global definition on $\cal N$ of 
the integrand. Hence, it is omitted at this stage. 

In view of applications to
concrete models, the discussion below is restricted to the family of 
equivariant $s$ cohomology classes $\cal O$ that are constant on the factors 
$\cal N$ and $\cal G$ of $\cal P$. These are of the form 
${\cal O}={\cal O}(a,\psi,\Omega)$ and thus represent 
equivariant $s$ cohomology classes of $\cal A$. 

In field theory, it is 
more natural to use instead of $g,~\epsilon$ the integration variables
$$
\chi=\ln g,\qquad\qquad\lambda=f(g)\zeta.
\eqno(3.47)
$$
This functional change of variables is not globally defined because 
the group logarithm $\ln$ is defined only near $g=1$ as mentioned earlier. 
However, in spite of this shortcoming, it is suitable for the field theoretic
analysis below. The topological action $S_{\rm top}$ given by (3.41)
can be expressed in terms of $\chi,~\lambda$. Since $\lambda=s\chi$ and the 
boson/fermion measures are invariant under functional changes of variables
with positive jacobian, one has that $dgd\epsilon=d\chi d\lambda$. 
So, denoting by $I_\alpha(n,\theta,a,\psi,\chi,\lambda,\omega,\Omega)$ the 
functional integral given by the equivalent expressions (3.40)--(3.41) and 
(3.44), we shall consider the functional integral 
$$
I_{{\cal O}\alpha}(n,\theta)=\int dad\psi\int d\chi d\lambda\int 
d\omega d\Omega I_\alpha(n,\theta,a,\psi,\chi,\lambda,\omega,\Omega)
{\cal O}(a,\psi,\Omega).
\eqno(3.48)
$$
It is now time to discuss the locality and renormalizability 
of the resulting quantum field theory.
\vskip .2cm
{\it Locality}. 
\vskip .2cm
The quantum field theory described by (3.48) is manifestly local provided
the functionals $H_\alpha(n,a,g)$ and $W(a,g)$ are judiciously chosen in 
such a way to yield local contributions to the action. In fact, 
if this is the case, the topological action 
$S_{\rm top}$ is a local functional of $g$ and $\zeta$
which, in turn, are given by local expressions in terms of $\chi$ and 
$\lambda$, namely $g=e^\chi$ and $\zeta=f(e^\chi)^{-1}\lambda$. Had one 
used as functional variables $g,~\epsilon$ instead of $\chi,~\lambda$, 
one would have had an action containing explicit occurrences of the non 
local functional $c^\dagger(g)$, spoiling locality.
\vskip .2cm
{\it Renormalizability}. 
\vskip .2cm
The quantum field theory yielded by (3.48) is not manifestly renormalizable
for two basic reasons. First, the action contains the non linear combination
$e^\chi$ which is generally incompatible with renormalizability. 
Second, the available Ward identities may not be sufficient 
to prevent the generation in the quantum theory of terms 
allowed by locality and power counting but not contained in the action. 
 
The following formal argument shows that the field theory considered here is 
equivalent in $s$ cohomology to one in which the group $\cal G$ is 
`flattened' into its Lie algebra $\Lie{\cal G}$. 
To begin with, one performs the following change of functional variables:
$$
\eqalign{&\chi' \cr 
         &r'_\alpha(n) \cr}
\eqalign{&\!\!=(1/k)\chi, \cr 
         &\!\!=(1/k)r_\alpha(n), \cr}
\qquad\qquad
\eqalign{&\lambda'\cr 
         &z'_\alpha(n)\cr}
\eqalign{&\!\!=(1/k)\lambda, \cr 
         &\!\!=(1/k)z_\alpha(n), \cr}
\eqno(3.49)
$$
where $k>0$. Since the boson/fermion measures are invariant under changes of 
coordinates with positive jacobian, one has that $d\chi' d\lambda'=d\chi 
d\lambda$ and $dr'_\alpha(n)dz'_\alpha(n)=dr_\alpha(n)dz_\alpha(n)$.
This being a mere change of variables, the resulting functional
integral is independent from $k$. Next, one rescales the metric 
$(\cdot,\cdot)$ of the $\bar\chi,~\bar\lambda$ and $u_\alpha(n),~v_\alpha(n)$ 
sectors into $(1/k')(\cdot,\cdot)$ and replaces $t$ by $t/k'$, where $k'>0$, 
leaving $s$ cohomology unchanged. 
Now, nothing forbids setting $k=k'$ and taking the limit 
$k=k'\to0$, provided the limit exists and is non singular.
Using (3.15)--(3.17) and the fact that $f(1)=1$, 
it is easy to see that $W(a,e^{k\chi'})=kD(a)\chi'+O(k^2)$ and that 
$\zeta=k\lambda'+O(k^2)$.
The limit affects the $\bar\tau,\bar\omega$, $\bar\chi,\bar\lambda$
and $u_\alpha(n),v_\alpha(n)$ sectors of the action (3.41), which now 
becomes 
$$
\eqalignno{
-S_{\rm flat}&=-{1\over4b}(\bar\tau,\bar\tau)_{\cal F}{}^\vee
-i\langle \bar\tau, H_\alpha(n,a,1)\rangle_{\cal F}
-i\langle\bar\omega,\iota_a(\psi+D(a)\omega)
-\theta\partial_nH_\alpha(n,a,1) &\cr
&\phantom{\simeq}\vphantom{\bigg\{}
-h_\alpha(n,a,1,\psi,-\omega)\rangle_{\cal F}
-{1\over4b}(\bar\omega,\Omega_{\cal F}{}^\vee\bar\omega)_{\cal F}{}^\vee
-{1\over4t}(\bar\chi,\bar\chi)-i(\bar\chi,D^\dagger D(a)\chi') &\cr
&\phantom{\simeq}\vphantom{\bigg\{}
+i(\bar\lambda,D^\dagger D(a)\lambda'
+(\psi+D(a)\omega)\partial D^\dagger(a)D(a)\chi' &\cr
&\phantom{\simeq}\vphantom{\bigg\{}
+D^\dagger(a)(T_2w(a,1)\chi')((\psi+D(a)\omega)\oplus 0))
-i(v_\alpha(n)+\theta\partial_nu_\alpha(n),\chi') &\cr
&\phantom{\simeq}\vphantom{\bigg\{}
+i(u_\alpha(n),\lambda')
+i(z'_\alpha(n)+\theta\partial_nr'_\alpha(n),\bar\lambda)
+i(r'_\alpha(n),\bar\chi)+\cdots, &(3.50) \cr}
$$
where $T_2w(a,1)$ is the tangent map of $w(a,g)$ with respect to $g$ at 
$g=1$ and the ellipses denote the remaining terms of the action unchanged 
in the limit. In this way, one eliminates the troublesome exponential
$e^\chi$ present in the topological action without changing the 
topological content of the theory.
It is remarkable that by this procedure the problem of the lack of global 
definition on $\cal G$ of the map $g\to\ln g$ is simultaneously cured. Note 
however that the localization sector of the action $S_{\rm flat}$ is still of 
the form (1.10) but with (1.11) no longer holding. This somewhat obscures the 
topological origin of $S_{\rm flat}$.

In refs. \ref{18,23}, it has been argued that, in order to ensure the 
renormalizability of the topological quantum field 
theory, the gauge fermion $\Phi_{FP}$
should be of the form $w\Psi_{FP}$, where $w$ is a nilpotent operation 
anticommuting with $s$ and annihilating all equivariant functionals such as 
$\Phi_{MQ}$ and $\Phi_{W}$ (see sect. 1). $w$ rather than $s$ would be 
the true counterpart of the BRST operator. 
To $w$ there are associated extra Ward identities which might be 
necessary for renormalizability. The gauge fermion which has been used above 
does not have this property. However, following the procedure of refs. 
\ref{18,23} would simply add to the Faddeev--Popov action used further terms 
depending on extra ghostly fields, which, in the functional integral of the 
quantum field theory considered here, would yield a trivial insertion $1$. 
While these terms may be necessary for the manifest renormalizability of the 
quantum field theory, they are not expected to affect 
its infrared properties which are the main object of our analysis. 

From the above discussion, it seems plausible that the topological quantum 
field theory constructed above is renormalizable, provided the functionals 
$H_\alpha(n,a,g)$ and $W(a,g)$ are properly chosen. Of course, these 
arguments can not be considered in any way a conclusive proof.
\vskip .2cm
{\it The regular irreducible stratum ${\cal B}^*$}. 
\vskip .2cm
Suppose that $\cal N$ is the stratum ${\cal B}^*$ of regular  
$\widehat{\cal G}$ orbits (see sect. 2). In this case, the topological action 
$S_{\rm top}$ and the flattened action $S_{\rm flat}$, given respectively 
by eqs. (3.41) and (3.50), simplify considerably. Indeed, since, for any 
$a\in{\cal A}^*$, $\widehat{\cal G}(a)=\widehat{\cal Z}$,
the auxiliary fields $\bar y_\alpha(n),~\bar x_\alpha(n)$, 
$x_\alpha(n),~y_\alpha(n)$, $u_\alpha(n),~v_\alpha(n)$,
$r_\alpha(n),~z_\alpha(n)$ or $r'_\alpha(n),~z'_\alpha(n)$ are valued 
in $\Lie{\cal Z}$ with ${\cal Z}={\cal G}\cap\widehat{\cal Z}$ and
so can be chosen independent from $\alpha$ and $n$. In this way, the 
only $n$ dependent object appearing in the actions is the 
local lift $a_\alpha(n)$. Moreover, if $\cal Z$ is trivial, 
the terms containing the auxiliary fields are identically zero. 
\vskip .4cm
\par\noindent
{\bf 4. The Gribov Problem and its Treatment}
\vskip .4cm
\par
The functional integral $I_{{\cal O}\alpha}(n,\theta)$ of eq. (3.48)
is defined on the local patch $U_\alpha$ of $\cal N$. Let us now find out
under which conditions $I_{{\cal O}\alpha}(n,\theta)$ is the local 
restriction on $U_\alpha$ of a globally defined functional
$I_{\cal O}(n,\theta)$ on $\cal N$. This is in essence the Gribov 
problem and amounts to checking whether 
$$
I_{\cal O\alpha}(n,\theta)=I_{\cal O\beta}(n,\theta)
\eqno(4.1)
$$
on $U_\beta\cap U_\alpha\not=\emptyset$.

The lifts $a_\alpha$ and $a_\beta$ of overlapping neighborhoods $U_\alpha$ 
and $U_\beta$ of $\cal N$ to $\cal A$ must satisfy a relation of the form
$$
a_\alpha(n)=a_\beta(n)\kappa_{\beta\alpha}(n)_{\cal A}
\eqno(4.2)
$$
on $U_\beta\cap U_\alpha$. The transition functions $\kappa_{\beta\alpha}
:U_\beta\cap U_\alpha\to\widehat{\cal G}$ are defined up to right 
multiplication by an element of $\widehat{\cal G}(a_\alpha(n))$ and left 
multiplication by an element of $\widehat{\cal G}(a_\beta(n))$. 
For this reason, in general, the $\widehat{\cal G}$ valued 
1--cochain $\kappa=\{\kappa_{\beta\alpha}\}$ is not 
a 1--cocycle and thus it does not define a principal $\widehat{\cal G}$ 
bundle over $\cal N$ \ref{34}. Instead of the 1--cocycle condition,
the $\kappa_{\beta\alpha}$ satisfy the more general condition 
$$
\kappa_{\gamma\beta}(n)\kappa_{\beta\alpha}(n)
=k_{\beta\alpha\gamma}(n)\kappa_{\gamma\alpha}(n)
\eqno(4.3)
$$
on any triple intersection $U_\beta\cap U_\alpha\cap U_\gamma\not=\emptyset$,
where $k_{\beta\alpha\gamma}(n)\in\widehat{\cal G}(a_\gamma(n))$.

The choice of the local lifts $a_\alpha:U_\alpha\to{\cal A}$ is 
conventional. If $a'_\alpha:U_\alpha\to{\cal A}$ is another choice
of local lifts, then there exist maps $\nu_\alpha:U_\alpha\to\widehat{\cal G}$ 
such that
$$
a'_\alpha(n)=a_\alpha(n)\nu_\alpha(n)_{\cal A}.
\eqno(4.4)
$$
With the lifts $a'_\alpha$, there is associated a new set of transition 
functions $\kappa'_{\beta\alpha}$ satisfying a relation of the same 
form as (4.3).

To carry out the Gribov analysis, one has to make some technical assumptions
verified in concrete models. First, the invariant subgroup $\widehat{\cal Z}$ 
of $\widehat{\cal G}$ acting trivially on $\cal A$ acts trivially on $\cal E$
as well:
$$
\gamma_{\cal E}=1_{\cal E},\qquad\qquad\gamma\in\widehat{\cal Z}.
\eqno(4.5)
$$
Second, the insertion ${\cal O}(a,\psi,\Omega)$ is flat in the 
direction of $\Lie{\cal Z}$, where ${\cal Z}={\cal G}\cap\widehat{\cal Z}$:
$$
{\cal O}(a,\psi,\Omega+\varepsilon\xi)={\cal O}(a,\psi,\Omega),
\qquad\qquad\xi\in\Lie{\cal Z},
\eqno(4.6)
$$
where $\varepsilon$ is a parameter of degree 2. 
One needs also a matching assumption on the localizing functional 
$H_\alpha(n,a,g)$, namely that 
$$
H_\beta(n,a,g)|_{g=1}=\kappa_{\beta\alpha}(n)_{\cal F} 
H_\alpha(n,a\kappa_{\beta\alpha}(n)_{\cal A},g)|_{g=1}
\eqno(4.7)
$$
on $U_\beta\cap U_\alpha\not=\emptyset$.

The Gribov analysis invokes repeatedly certain matching relations collected 
below. Using the definition of $H^*_\alpha(n,a,g)$ given below (3.9), the 
properties of the map $\iota$ listed below eq. (3.5) and (4.7), it is 
straightforward to show that
$$
H^*_\alpha(n,a,1)
=TR_{{\cal A}\kappa_{\beta\alpha}(n)}
(a\kappa_{\beta\alpha}(n)^{-1}{}_{\cal A})
H^*_\beta(n,a\kappa_{\beta\alpha}(n)^{-1}{}_{\cal A},1)
\eqno(4.8)
$$
on $U_\beta\cap U_\alpha\not=\emptyset$.
Using (4.2) and the basic identities $D^\dagger D(a\gamma_{\cal A})
=\Ad\gamma^{-1}D^\dagger D(a)\Ad\gamma$ and $D(a\gamma_{\cal A})=
TR_{{\cal A}\gamma}(a)D(a)\Ad\gamma$ for $\gamma\in\widehat{\cal G}$
and recalling that $q(a\gamma_{\cal A})
=\Ad\gamma^{-1}q(a)\Ad\gamma$, the following relations are easily obtained:
$$
\theta\partial_na_\alpha(n)=
TR_{{\cal A}\kappa_{\beta\alpha}(n)}(a_\beta(n))\theta\partial_na_\beta(n)
+D(a_\alpha(n))(\kappa_{\beta\alpha}(n)^{-1}
\theta\partial_n\kappa_{\beta\alpha}(n))
\eqno(4.9)
$$
and 
$$
\upsilon_\alpha(n)
=\kappa_{\beta\alpha}(n)^{-1}
\theta\partial_n\kappa_{\beta\alpha}(n)
+\Ad\kappa_{\beta\alpha}(n)^{-1}\upsilon_\beta(n)
-\sigma_{\beta\alpha}(n),
\eqno(4.10)
$$
$$
\Upsilon_\alpha(n)=\Ad\kappa_{\beta\alpha}(n)^{-1}\Upsilon_\beta(n)
-\Sigma_{\beta\alpha}(n),
\eqno(4.11)
$$
where 
$$
\sigma_{\beta\alpha}(n)=
q(a_\alpha(n))(\kappa_{\beta\alpha}(n)^{-1}
\theta\partial_n\kappa_{\beta\alpha}(n)),
\eqno(4.12)
$$
$$
\Sigma_{\beta\alpha}(n)=
(\theta\partial_n+\ad\upsilon_\alpha(n))\sigma_{\beta\alpha}(n)
+(1/2)[\sigma_{\beta\alpha}(n),\sigma_{\beta\alpha}(n)],
\eqno(4.13)
$$
on $U_\beta\cap U_\alpha\not=\emptyset$. 

Now, from (3.44), it is clear that
$$
I_\alpha(n,\theta,a,\psi,\chi,\lambda,\omega,\Omega)
=I(a,\psi,\chi,\lambda,\omega,\Omega;a_\alpha(n),\theta\partial_na_\alpha(n),
\upsilon_\alpha(n),\Upsilon_\alpha(n)).
\eqno(4.14)
$$
Using (3.2), (4.8)--(4.11) and the $\widehat{\cal G}$ invariance of the 
insertion ${\cal O}(a,\psi,\Omega)$ and recalling the formal rules stated in
footnote 5, it is straightforward to show that
$$\eqalignno{
&I(a,\psi,\chi,\lambda,\omega,\Omega;a_\alpha(n),\theta\partial_na_\alpha(n),
\upsilon_\alpha(n),\Upsilon_\alpha(n)){\cal O}(a,\psi,\Omega) &\cr
=~&I(a\kappa_{\beta\alpha}(n)^{-1}{}_{\cal A},
\psi\kappa_{\beta\alpha}(n)^{-1}{}_{\cal A},\chi,\lambda,
\Ad\kappa_{\beta\alpha}(n)(\omega-\kappa_{\beta\alpha}(n)^{-1}
\theta\partial_n\kappa_{\beta\alpha}(n)+\sigma_{\beta\alpha}(n)), &\cr
\phantom{=~}&
\Ad\kappa_{\beta\alpha}(n)\Omega;a_\beta(n),\theta\partial_na_\beta(n),
\upsilon_\beta(n),\Upsilon_\beta(n)
-\Ad\kappa_{\beta\alpha}(n)\Sigma_{\beta\alpha}(n)) &\cr
\phantom{=~}&
\times{\cal O}(a\kappa_{\beta\alpha}(n)^{-1}{}_{\cal A},
\psi\kappa_{\beta\alpha}(n)^{-1}{}_{\cal A},
\Ad\kappa_{\beta\alpha}(n)\Omega) &\cr
\phantom{=~}&
\times\sgn\dtr(\kappa_{\beta\alpha}(n)_{\cal E})
\sgn\dtr(\kappa_{\beta\alpha}(n)^{-1}{}_{\cal A})
\sgn\dtr(\Ad\kappa_{\beta\alpha}(n)).&(4.15)\cr}
$$
From here, it follows immediately that
$$\eqalignno{
&\int dad\psi\int d\chi d\lambda\int d\omega d\Omega 
I(a,\psi,\chi,\lambda,\omega,\Omega;a_\alpha(n),\theta\partial_na_\alpha(n),
\upsilon_\alpha(n),\Upsilon_\alpha(n)){\cal O}(a,\psi,\Omega) &\cr
=~&\sgn\dtr(\kappa_{\beta\alpha}(n)_{\cal E})
\int dad\psi\int d\chi d\lambda\int d\omega d\Omega 
I(a,\psi,\chi,\lambda,\omega,\Omega;a_\beta(n),\theta\partial_na_\beta(n),
\upsilon_\beta(n), &\cr
\phantom{=~}&\vphantom{\int}\Upsilon_\beta(n)
-\Ad\kappa_{\beta\alpha}(n)\Sigma_{\beta\alpha}(n))
{\cal O}(a,\psi,\Omega), &(4.16)\cr}
$$
since $d(a\gamma^{-1}{}_{\cal A})
d(\psi\gamma^{-1}{}_{\cal A})=
\sgn\dtr(\gamma^{-1}{}_{\cal A})dad\psi$ and 
$d(\Ad\gamma \omega)d(\Ad\gamma \Omega)
=\sgn\dtr(\Ad\gamma )$ $d\omega d\Omega$ for $\gamma\in\widehat{\cal G}$.
From (3.48), (4.14) and (4.16), it appears that (4.1) is not fulfilled  
in general. A possible mechanism 
by which the Gribov ambiguity may nevertheless be solved is the following. 

Let $q_{\Lie{\cal Z}}$ be the orthogonal projector of $\Lie{\cal G}$
onto $\Lie{\cal Z}$. Let assume that
$$
q_{\Lie{\cal Z}}\Sigma_{\beta\alpha}(n)=\Sigma_{\beta\alpha}(n).
\eqno(4.17)
$$
Then, recalling that $I_\alpha(n,\theta,a,\psi,\chi,\lambda,\omega,\Omega)$ 
is the functional integral given by (3.44) and that (4.14) holds,
it is is easy to show using (4.5), (4.6) and (4.17) that
$$
\eqalignno{
&I(a,\psi,\chi,\lambda,\omega,\Omega;a_\beta(n),\theta\partial_na_\beta(n),
\upsilon_\beta(n),\Upsilon_\beta(n)
-\Ad\kappa_{\beta\alpha}(n)\Sigma_{\beta\alpha}(n))
{\cal O}(a,\psi,\Omega), &\cr
=~&I(a,\psi,\chi,\lambda,\omega,
\Omega+\Ad\kappa_{\beta\alpha}(n)\Sigma_{\beta\alpha}(n);
a_\beta(n),\theta\partial_na_\beta(n),\upsilon_\beta(n),\Upsilon_\beta(n)) &\cr
\phantom{=~}&\times
{\cal O}(a,\psi,\Omega+\Ad\kappa_{\beta\alpha}(n)\Sigma_{\beta\alpha}(n)). 
&(4.18) \cr}
$$
Substituting this relation in (4.16) and taking into account (3.48) and 
(4.14), one finds that
$$
I_{\cal O\alpha}(n,\theta)=s_{\beta\alpha}I_{\cal O\beta}(n,\theta)
\eqno(4.19)
$$ 
on $U_\beta\cap U_\alpha\not=\emptyset$, where
$$
s_{\beta\alpha}=\sgn\dtr(\kappa_{\beta\alpha}(n)_{\cal E}).
\eqno(4.20)
$$
$s_{\beta\alpha}=\pm1$ is independent from $n$ if $U_\beta\cap U_\alpha$ 
is connected, as is assumed for simplicity. (4.19) is weaker than (4.1) 
because the sign ambiguity associated with $s_{\beta\alpha}$. We shall come 
back to this point shortly.

In general, (4.17) does not hold. However, one may change the local lifts 
according to (4.4) and impose that the $\Sigma'_{\beta\alpha}$
satisfy (4.17). Using (4.4), (4.12) and (4.13), recalling that 
$\kappa'_{\beta\alpha}(n)=\nu_\beta(n)^{-1}\kappa_{\beta\alpha}(n)
\nu_\alpha(n)$ and going through steps analogous to those involved in the 
derivation of (4.10), it is straightforward to show that this amounts to 
solving the equation 
$$
\eqalignno{
\Sigma'_{\beta\alpha}(n)=&\Ad\nu_\alpha(n)^{-1}\big[\Sigma_{\beta\alpha}(n)-
(\theta\partial_n+\ad\upsilon_\alpha(n))\chi_\alpha(n)
-(1/2)[\chi_\alpha(n),\chi_\alpha(n)] &\cr
&+\Ad\kappa_{\beta\alpha}(n)^{-1}
(\theta\partial_n+\ad\upsilon_\beta(n))\chi_\beta(n)
+(1/2)[\chi_\beta(n),\chi_\beta(n)])\big], &\cr
\chi_\alpha(n)
=&-q(a_\alpha(n))(\theta\partial_n\nu_\alpha(n)\nu_\alpha(n)^{-1})
\vphantom{\big[} &(4.21)\cr}
$$
for the $\nu_\alpha$ with $\Sigma'_{\beta\alpha}$ fulfilling (4.17).
Before posing the question of the existence of a solution of eq. (4.21),
one has to check whether it is consistent with (4.3) and with the matching 
relation (4.11).

From (4.3) and (4.11), one has
$$
\Sigma_{\gamma\alpha}(n)-\Sigma_{\beta\alpha}(n)
-\Ad\kappa_{\beta\alpha}(n)^{-1}\Sigma_{\gamma\beta}(n)
=-\Ad\kappa_{\gamma\alpha}(n)^{-1}(\Ad k_{\beta\alpha\gamma}(n)^{-1}-1)
\Upsilon_\gamma(n).
\eqno(4.22)
$$
If $\Sigma_{\beta\alpha}$ satisfies (4.17), then the left hand side of 
the above equation is valued in $\Lie{\cal Z}$, as $\Lie{\cal Z}$ is
invariant under the adjoint action of $\widehat{\cal G}$. So must be the 
right hand side. It seems unlikely that this can come about unless 
the $k_{\beta\alpha\gamma}$ are $\widehat{\cal Z}$ valued. This conclusion 
remains unchanged even if $\Sigma_{\beta\alpha}$ does not satisfy (4.17),
but eq. (4.21) can be solved, since, under the replacement (4.4),
$k'_{\beta\alpha\gamma}(n)=\nu_\gamma(n)^{-1}
k_{\beta\alpha\gamma}(n)\nu_\gamma(n)$ and $\widehat{\cal Z}$
is an invariant subgroup of $\widehat{\cal G}$. 

From now on, we shall thus assume that the $k_{\beta\alpha\gamma}$ are 
$\widehat{\cal Z}$ valued. If this is not the case,
the analysis below can still be carried out, but it becomes much 
messier and has no apparent sheaf theoretic interpretation. 

From (4.3), it follows then that the 
$\widehat{\cal G}$ valued 1--cochain $\kappa$ defines a 
$\widehat{\cal G}/\widehat{\cal Z}$ valued 1--cocycle $\bar\kappa$.
Eq. (4.22) states then $\Sigma=\{\Sigma_{\beta\alpha}\}$ is a 1--cocycle of
$\bigwedge^2T^*{\cal N}\otimes\Ad\bar\kappa$, which, on account of (4.11),
is trivial \ref{34}. This, however, is not sufficient to
guarantee the solvability of eq. (4.21) for the $\nu_\alpha$. 

The $\widehat{\cal G}/\widehat{\cal Z}$ valued 1--cocycle $\bar\kappa$
defines a principal $\widehat{\cal G}/\widehat{\cal Z}$ bundle 
${\cal A}_{\cal N}$ on $\cal N$. The existence of a solution 
$\nu=\{\nu_\alpha\}$ of eq. (4.21) is a condition much stronger than 
the simple triviality of the 1--cocycle $\Sigma$; it is strongly
reminiscent of a flatness condition and may entail topological 
restrictions for ${\cal A}_{\cal N}$. Recall that, in finite dimension, for 
any principal $G$ bundle $P$ on $X$ with transition functions $g_{ba}$, the 
local $\goth g$ valued 1--forms $g_{ba}{}^{-1}dg_{ba}$ always define
a trivial 1--cocycle of $\bigwedge^1T^*X\otimes\Ad g$, a fact indeed
equivalent to the existence of a connection on $P$. So, 
$g_{ba}{}^{-1}dg_{ba}=m_a-\Ad g_{ba}{}^{-1}m_b$ for certain local 
$\goth g$ valued 1--forms $m_c$. In order the bundle $P$ to be flat, 
it is necessary and sufficient that the $m_c$ can be chosen of the form 
$m_c=-dg_cg_c{}^{-1}$ for certain local $G$ valued function $g_c$. This is 
a non trivial requirement with topological implications for $P$. 

Let us come to the problem of the sign $s_{\beta\alpha}$ in (4.19). 
With the action of $\widehat{\cal G}$ on $\cal E$ there is associated 
the vector bundle ${\cal V}_{\cal N}={\cal A}_{\cal N}
\times_{\widehat{\cal G}/\widehat{\cal Z}}{\cal E}$ on $\cal N$. 
the transition functions of such a bundle are precisely the 
$\kappa_{\beta\alpha{\cal E}}$. From the definition of 
$s_{\beta\alpha}$, eq. (4.20), it appears that requiring that 
$s_{\beta\alpha}=1$
is tantamount to demanding that ${\cal V}_{\cal N}$ is oriented. 

From the above discussion, we draw the following conclusions. 
{\it The Gribov problem is solvable provided the topology
of the principal $\widehat{\cal G}/\widehat{\cal Z}$ bundle 
${\cal A}_{\cal N}$ allows for the solution of eq. (4.21)
and the associated vector bundle ${\cal V}_{\cal N}$ is oriented}.

Let us consider the important case where $\cal N$ is the stratum ${\cal B}^*$ 
of regular $\widehat{\cal G}$ orbit. In this case, the $k_{\beta\alpha\gamma}$
are necessarily valued in $\widehat{\cal Z}$, as $\widehat{\cal G}(a)
=\widehat{\cal Z}$ for any irreducible element $a\in{\cal A}^*$, so that
our basic assumption is fulfilled. The principal $\widehat{\cal G}/
\widehat{\cal Z}$ bundle ${\cal A}_{\cal N}$ associated with 
$\kappa^*\equiv\bar\kappa$ may be identified with ${\cal A}^*$ \ref{32}. 
Correspondingly, the vector bundle ${\cal V}_{\cal N}$ is 
${\cal V}^*={\cal A}^*\times_{\widehat{\cal G}/\widehat{\cal Z}}{\cal E}$ 
\ref{32}.

For ${\cal N}={\cal B}^*$, (4.17) is automatically satisfied. 
This follows from (4.12)--(4.13), the fact that 
$q(a)=q_{\Lie{\cal Z}}$ for any irreducible $a\in{\cal A}^*$ and 
the invariance of the Lie subalgebra $\Lie{\cal G}$.
Therefore, there is no need to solve eq. (4.21)!
This shows that {\it  for the regular 
stratum ${\cal B}^*$ of the $\widehat{\cal G}$ orbit space $\cal B$,
the Gribov problem is solved, provided ${\cal V}^*$ is oriented}.

When ${\cal N}\not={\cal B}^*$ very little can be said. Though we do 
not have conclusive evidence, it seems unlikely that for a singular
stratum $\cal N$ eq. (4.21) is consistent with (4.22). 

When the Gribov problem is solvable for the stratum $\cal N$ of $\cal B$, 
the map ${\cal O}\to I_{\cal O}$ defines a homomorphism of the equivariant 
cohomology $H^*_{\rm equiv}({\cal A})$ of $\cal A$ into the cohomology 
$H^*({\cal N})$ of $\cal N$. This is  not difficult to show. 
Let $X$ denote the collection of all fields but $n,~\theta$.
For any insertion ${\cal K}(X)$, the relevant functional integrals
are of the form
$$
I_{\cal K}(n,\theta)=\int dX e^{s\Psi(n,\theta,X)}{\cal K}(X),
\eqno(4.23)
$$
where $\Psi(n,\theta,X)$ is the appropriate gauge fermion.
Then, denoting by $s|_X$ the restriction of $s$ to the fields $X$, one has 
$$
\eqalignno{
\theta\partial_nI_{\cal K}(n,\theta)&=
\int dX (s-s|_X)e^{s\Psi(n,\theta,X)}{\cal K}(X) &\cr
&=\int dX e^{s\Psi(n,\theta,X)}s|_X{\cal K}(X)
-\int dXs|_X\big( e^{s\Psi(n,\theta,X)}{\cal K}(X)\big) &\cr
&=\int dX e^{s\Psi(n,\theta,X)}s{\cal K}(X). &(4.24)\cr}
$$
This relation shows that $I_{\cal K}(n,\theta)$ is $\theta\partial_n$ closed
(exact) whenever $\cal K$ is $s$ closed (exact).

Assume that $\cal N$ has no boundary to have a sensible intersection 
theory. For a given ${\cal O}\in H^*_{\rm equiv}({\cal A})$, its image  
$I_{\cal O}\in H^*({\cal N})$ factorizes as
$$
I_{\cal O}(n,\theta)=PD_{{\cal N}_F}(n,\theta)\cdot J_{\cal O}(n,\theta).
\eqno(4.25)
$$
Here, $PD_{{\cal N}_F}$ is formally the Poincar\'e dual of the submanifold 
${\cal N}_F$ of $\cal N$ defined by the equivariant condition 
$F(a)=0$ and is independent from $\cal O$ and $J_{\cal O}$ is an element 
of $H^*({\cal N})$.  This is fairly evident from the local 
expression (3.48) of $I_{\cal O}(n,\theta)$ and from the structure of the 
integrand $I_\alpha(n,\theta,a,\psi,\chi,\lambda,\omega,\Omega)$ given in 
(3.44). Indeed, the Mathai--Quillen $\varrho,~\varpi$ integral is just a 
regularized version of the boson/fermion $\delta$ function pair
$\delta(F(a))\delta(\psi\partial F(a))$ \ref{23}. In concrete models
${\cal N}_F$ is finite dimensional and $J_{\cal O}$ has finite degree
in $H^*({\cal N})$. Were one able to define integration on 
$\cal N$, the above formula would be the starting point for the study of the 
intersection theory of ${\cal N}_F$ using field theory. This is however a 
technically non trivial problem, since $\cal N$ is generally infinite 
dimensional.
\vskip .4cm
\par\noindent
{\bf 5. The Donaldson--Witten Model}
\vskip .4cm
\par
In this section, we shall apply the formalism developed in the previous 
section to the Donaldson Witten model as an illustration. 

Let $X$ be an oriented connected compact 4--manifold equipped with a metric 
$h$. Let $G$ be a reductive compact Lie group. Finally, let $B\to X$ be 
a principal $G$ bundle over $X$. Consider the space 
$$
{\cal A}=\Conn(B)
\eqno(5.1)
$$
of connections of $B$. $\cal A$ is an affine space modelled on 
$\Omega^1(X,\Ad B)$. Hence, for any $a\in{\cal A}$, one has the 
canonical identification $T_a{\cal A}\simeq\Omega^1(X,\Ad B)$. Thus,
$$
\psi\in\Pi_1\Omega^1(X,\Ad B)
\eqno(5.2)
$$
\footnote{}{}
\footnote{${}^6$}{
$\Pi_r\Omega^p(X,\Ad B)$ denotes the space of $\Ad B$ valued $p$ forms
of Grassman degree $r$.}. $T_a{\cal A}$ carries a natural metric, 
the standard metric on $\Omega^1(X,\Ad B)$ associated with $h$ and 
a negative definite $\Ad G$ invariant extension $\Tr$ of the Cartan-Killing 
form $\tr$ of $\goth g$.
\footnote{}{}
\footnote{${}^7$}{For any $p$, this metric is given by $(\alpha,\beta)=
-\int_X\Tr(\alpha\wedge*\beta)$ for $\alpha,~\beta\in\Omega^p(X,\Ad B)$. 
$*$ is the Hodge star operator associated with $h$.
$\Tr$ may be constructed as follows. Let $\goth z$ 
be the center of $\goth g$. Let $<\cdot,\cdot>$ be a negative 
definite symmetric bilinear form on $\goth z$.
Now, $\goth g=\goth g/\goth z\oplus\goth z$ with
$\goth g/\goth z$ semisimple. Correspondingly, every $x\in\goth g$ decomposes
uniquely as $x'+x_0$, where $x'\in\goth g/\goth z$ and $x_0\in\goth z$. 
Then, for $x,~y \in\goth g$, $\Tr(xy)=\tr(x'y')+<x_0,y_0>$. $\Tr$ is 
manifestly $\Ad G$ invariant. If $G$ has a discrete center $Z$, 
then $\goth g$ is semisimple and $\Tr$ reduces to the usual Cartan--Killing
form.}.

Consider the gauge group
$$
\widehat{\cal G}=\Gau(B)
\eqno(5.3)
$$
of the principal $G$ bundle $B$ and the subgroup $\cal G$ of
$\widehat{\cal G}$ of the elements homotopic to the identity.
One has the canonical isomorphism $\Lie{\cal G}\simeq\Omega^0(X,\Ad B)$. 
Thus, 
$$
\zeta\in\Pi_1\Omega^0(X,\Ad B).
\eqno(5.4)
$$
$\Lie{\cal G}$ carries the standard metric on $\Omega^0(X,\Ad B)$ 
associated with $h$ and $\Tr$, which is $\Ad\widehat{\cal G}$ invariant. 
To this there is associated a metric on $\cal G$ in standard fashion 
(cfr. footnote 4 above). 

$\widehat{\cal G}$ acts on $\cal A$ by gauge transformations: for $a\in
{\cal A}$ and $\gamma\in\widehat{\cal G}$, $a\gamma_{\cal A}=
\gamma^{-1}d\gamma+\Ad\gamma^{-1}a$.  
The vertical vector fields of $\cal A$ are of the form $D(a)\xi$ for 
$\xi\in\Lie{\cal G}$, where $D(a)$ is the usual gauge covariant
derivative of the connection $a\in{\cal A}$: $D(a)\xi=d\xi+[a,\xi]$.

The subgroup $\widehat{\cal Z}$ of 
$\widehat{\cal G}$ acting trivially on $\cal A$ consists of the constant
elements of $\widehat{\cal G}$ valued in the center $Z$ of $G$,
so that $\widehat{\cal Z}\simeq Z$.
Correspondingly, the Lie algebra $\Lie{\cal Z}$ 
of ${\cal Z}=\widehat{\cal Z}\cap{\cal G}$
consists of the constant elements of $\Lie{\cal G}$ valued in 
the center $\goth z$ of $\goth g$ and  $\Lie{\cal Z}\simeq{\goth z}$. 
$\Lie{\cal Z}$
is clearly contained in the kernel of $D(a)$ for any $a\in{\cal A}$. 

Since $\Lie{\cal G}\simeq\Omega^0(X,\Ad B)$, one has 
$$
\eqalign{
&\omega\in\Pi_1\Omega^0(X,\Ad B), \cr
&\Omega\in\Pi_2\Omega^0(X,\Ad B). \cr}
\eqno (5.5)
$$
\vskip .2cm
{\it The Mathai--Quillen localization sector}. 
\vskip .2cm
The localization sector 
consists of two subsectors based on the infinite dimensional vector spaces
$$
\eqalignno{
{\cal E}&=\Omega^2_-(X,\Ad B),&(5.6)\cr
{\cal F}&=\Omega^1(X,\Ad B).&(5.7)\cr}
$$
$\cal E$ and $\cal F$ carry the standard metrics associated with $h$ and $\Tr$
and the standard adjoint action of $\widehat{\cal G}$. 
The localization functions are the antiselfdual part of the curvature
$F_-:{\cal A}\to{\cal E}$ 
$$
F_-(a)=(da+(1/2)[a,a])_-,
\eqno(5.8)
$$
as usual, and the map 
$H_\alpha:U_\alpha\times{\cal A}\times{\cal G}\to{\cal F}$
$$
H_\alpha(n,a,g)=\Ad g^{-1}(D(a)gg^{-1}-a+a_\alpha(n)),
\eqno(5.9)
$$
where $a_\alpha(n)$ is a local gauge slice on $U_\alpha$.
It is a straightforward matter to check that
$F_-(a)$ satisfies (3.2) and that $H_\alpha(n,a,g)$ satisfies (3.3), 
(3.4)--(3.5), (3.6)--(3.7) and (4.7). ($\iota_a$ is just the isomorphism
$T_a{\cal A}\simeq\Omega^1(X,\Ad B)={\cal F}$.) The Mathai--Quillen fields 
are
$$
\eqalign{
&\varrho\in\Pi_{-1}\Omega_-^2(X,\Ad B), \cr
&\varpi\in\Pi_0\Omega_-^2(X,\Ad B), \cr
&\bar\omega\in\Pi_{-1}\Omega^3(X,\Ad B), \cr
&\bar\tau\in\Pi_0\Omega^3(X,\Ad B). \cr}
\eqno(5.10)
$$
\vskip .2cm
{\it The Weil projection sector}.
\vskip .2cm
The Weil projection sector is identical to that of the customary 
Donaldson theory. The Weil fields are 
$$
\eqalign{
&\bar\Omega\in\Pi_{-2}\Omega^0(X,\Ad B), \cr
&\bar\psi\hskip 1pt\in\Pi_{-1}\Omega^0(X,\Ad B). \cr}
\eqno (5.11)
$$
\vskip .2cm
{\it The Faddeev--Popov gauge fixing sector}.
\vskip .2cm
The gauge fixing function has the general form
$$
\Sigma(a,g)=D^\dagger(a)W(a,g)+\mu^2\ln g,
\eqno(5.12)
$$
where $W(a,g)$ is chosen to be
$$
W(a,g)=D(a)gg^{-1}.
\eqno(5.13)
$$
By an infinitesimal argument, it is easy to see that for $\mu^2>0$,
the constraint $\Sigma(a,g)=0$ implies $g=1$ at least for $g$ near $1$.
It is conceivable that solutions other than $g=1$ exist. This would 
yield non perturbative effects that remain to be explored.
It is simple to check that $W(a,g)$ satisfies (3.15)--(3.17).
The Faddeev--Popov fields are 
$$
\eqalign{
&\bar\lambda\in\Pi_{-1}\Omega^0(X,\Ad B), \cr
&\bar\chi\in\Pi_0\Omega^0(X,\Ad B). \cr}
\eqno(5.14)
$$
\vskip .2cm
{\it The zero mode sector.}
\vskip .2cm
The field content of the zero mode sector is 
$$
\eqalign{
&\bar y(n) \in\Pi_{-2}\Omega^0(X,\Ad B), \cr
&\bar x(n)\in\Pi_{-1}\Omega^0(X,\Ad B), \cr
&x(n)\in\Pi_1\Omega^0(X,\Ad B), \cr
&y(n)\in\Pi_2\Omega^0(X,\Ad B), \cr
&u(n)\in\Pi_{-1}\Omega^0(X,\Ad B), \cr
&v(n)\in\Pi_0\Omega^0(X,\Ad B), \cr
&r(n)\in\Pi_0\Omega^0(X,\Ad B), \cr
&z(n)\in\Pi_1\Omega^0(X,\Ad B), \cr}
\qquad\qquad
\eqalign{
&D(a(n))\bar y(n)=0, \cr
&D(a(n))\bar x(n)=0, \cr
&D(a(n))x(n)=0, \cr
&D(a(n))y(n)=0, \cr
&D(a(n))u(n)=0, \cr
&D(a(n))v(n)=0, \cr
&D(a(n))r(n)=0, \cr
&D(a(n))z(n)=0. \cr}
\eqno(5.15)
$$
\vskip .2cm
{\it The Lagrangian.}
\vskip .2cm
From (3.43), the topological action $S_{\rm top}$ 
of the Donaldson-Witten model is given by
$$
\eqalignno{
S_{\rm top}&=-{1\over4}\int_X\Tr(\varpi\wedge*\varpi)
-i\int_X\Tr(\varpi\wedge F_-(a))-i\int_X\Tr(\varrho\wedge(D(a){\psi})_-) &\cr
&\phantom{=}
+{1\over4}\int_X\Tr(\varrho\wedge*[\Omega,\varrho])
-{1\over4b}\int_X\Tr(\bar\tau\wedge*\bar\tau)
-i\int_X\Tr(\bar\tau\wedge\Ad g^{-1}(D(a)gg^{-1} &\cr
&\phantom{=}-a+a(n)))
-i\int_X\Tr(\bar\omega\wedge(D(a)\zeta+\psi-D(a)\omega
-\Ad g^{-1}\theta\partial_na(n) &\cr
&\phantom{=}
+[\zeta-\omega,\Ad g^{-1}(D(a)gg^{-1}-a +a(n))]))
-{1\over4b}\int_X\Tr(\bar\omega\wedge*[\Omega,\bar\omega]) &\cr
&\phantom{=}
-i\int_X\Tr(D(a)\bar\psi\wedge*\psi)
+i\int_X\Tr(D(a)\Omega\wedge* D(a)\bar\Omega)
+i\int_X\Tr(\psi\wedge*[\bar\Omega,\psi]) &\cr
&\phantom{=}-{1\over{4t}}\int_X\Tr(\bar\chi\wedge*\bar\chi) 
-i\int_X\Tr((D(a)\bar\chi+[\bar\lambda,\psi-D(a)\omega])\wedge*D(a)gg^{-1}) 
&\cr
&\phantom{=}
-i\int_X\Tr(D(a)\bar\lambda\wedge*(D(a)\zeta
+(\Ad g-1)(D(a)\zeta+\psi-D(a)\omega))) &\cr
&\phantom{=}
-i\int_X\Tr((y(n)+\theta\partial_nx(n))\wedge*\bar\Omega)
+i\int_X\Tr(x(n)\wedge*(\bar\psi-[\omega,\bar\Omega])) &\cr
&\phantom{=}+i\int_X\Tr((\bar x(n)+\theta\partial_n\bar y(n))\wedge*\omega)
+i\int_X\Tr(\bar y(n)\wedge*(\Omega-(1/2)[\omega,\omega])) &\cr
&\phantom{=}-i\int_X\Tr((v(n)+\theta\partial_n u(n))\wedge*\ln g)
+i\int_X\Tr(u(n)\wedge*f(g)\zeta) &\cr
&\phantom{=}+i\int_X\Tr((z(n)+\theta\partial_nr(n))\wedge*\bar\lambda)
+i\int_X\Tr(r(n)\wedge*\bar\chi). &(5.16) \cr}
$$
To write the renormalizable action, one needs to introduce the rescaled fields 
$\chi',~\lambda'$, defined by (3.47) and (3,49), and the rescaled fields
$r'(n),~z'(n)$. Clearly,
$$
\eqalign{
&\chi'\in\Pi_0\Omega^0(X,\Ad B), \cr
&\lambda'\in\Pi_1\Omega^0(X,\Ad B), \cr}
\eqno(5.17)
$$
whereas $r'(n),~z'(n)$ are of the same type as $r(n),~z(n)$.
Then, by (3.50), the flattened action $S_{\rm flat}$
of the Donaldson--Witten model reads
$$
\eqalignno{
S_{\rm flat}&=
-{1\over4b}\int_X\Tr(\bar\tau\wedge*\bar\tau)
+i\int_X\Tr(\bar\tau\wedge(a-a(n)))
+i\int_X\Tr(\bar\omega\wedge(D(a)\omega-\psi &\cr
&\phantom{=}
+\theta\partial_na(n) 
-[\omega,a-a(n)]))
-{1\over4b}\int_X\Tr(\bar\omega\wedge*[\Omega,\bar\omega])
-{1\over{4t}}\int_X\Tr(\bar\chi\wedge*\bar\chi)  &\cr
&\phantom{=}
-i\int_X\Tr((D(a)\bar\chi
+[\bar\lambda,\psi-D(a)\omega])\wedge*D(a)\chi')
-i\int_X\Tr(D(a)\bar\lambda\wedge*(D(a)\lambda' &\cr
&\phantom{=}
+[\chi',\psi-D(a)\omega]))
-i\int_X\Tr((v(n)+\theta\partial_n u(n))\wedge*\chi')
+i\int_X\Tr(u(n)\wedge*\lambda') &\cr
&\phantom{=}
+i\int_X\Tr((z'(n)+\theta\partial_nr'(n))\wedge*\bar\lambda)
+i\int_X\Tr(r'(n)\wedge*\bar\chi)+\cdots, &(5.18)\cr}
$$
where the ellipses denote the remaining terms of $S_{\rm flat}$
which are the same as the corresponding terms of $S_{\rm top}$.

The topological observables of the Donaldson--Witten model, obtained from 
the descent equation \ref{12}, are given by the well-known expressions
$$
\eqalignno{
{\cal O}_0(a,\psi,\Omega)&={1\over 8\pi^2}\int_{C_0}\tr(\Omega^2), &\cr
{\cal O}_1(a,\psi,\Omega)&={1\over 8\pi^2}\int_{C_1}\tr(2\Omega\psi), &\cr
{\cal O}_2(a,\psi,\Omega)&={1\over 8\pi^2}\int_{C_2}\tr(2\Omega F(a)
+\psi\wedge\psi), &\cr
{\cal O}_3(a,\psi,\Omega)&={1\over 8\pi^2}\int_{C_3}\tr(2\psi\wedge F(a)), 
&\cr
{\cal O}_4(a,\psi,\Omega)&={1\over 8\pi^2}\int_{C_4}\tr(F(a)\wedge F(a)), 
&(5.19) \cr}
$$
where $C_k$ is a $k$--cycle of $X$. (Recall that $C_4=X$ and that $C_0$ 
is a finite set of points of $X$.)
Note that here the true Cartan--Killing form $\tr$ rather than its extension 
$\Tr$ appears.

The basic assumptions of the Gribov analysis of sect. 4 are fulfilled  in the 
present model. (4.5) is obviously satisfied,
since $\widehat{\cal G}$ acts on $\cal E$ by the adjoint action and this 
is trivial when restricted to the $Z$ valued elements of $\widehat{\cal Z}$.
(4.6) is satisfied, as appears from (5.19) by inspection, observing 
that the Cartan--Killing form $\tr$ vanishes on  
the $\goth z$ valued elements of $\Lie{\cal Z}$. (4.7) is also
satisfied as noticed earlier in this section.
So, from the analysis of sect. 4, it follows that 
{\it not only for the standard case 
$G=SU(2),~SO(3)$ but also for a general compact group $G$ 
the Gribov problem is solvable
for the regular irreducible stratum ${\cal B}^*$, provided 
${\cal V}^*={\cal A}^*\times_{\widehat{\cal G}/\widehat{\cal Z}}{\cal E}$
is oriented}. This conclusion, albeit based on formal manipulations of
functional integrals rather than on rigorous mathematics, is the main result
of this paper. 
\vskip .4cm
\par\noindent
{\bf 6. Concluding Remarks}
\vskip .4cm
\par
The real challenge of cohomological topological 
field theory is expressing intersection theory on moduli spaces in 
the language of local renormalizable field theory. This is not an easy 
task. While it is relatively easy to cook up local topological actions,
showing their renormalizability is a non trivial problem. The point is that 
locality and renormalizability are essentially field theoretic notions 
to which geometry may be quite indifferent. Our experience 
in gauge theories has taught us that it is difficult carry out gauge 
fixing salvaging renormalizability at the same time. However, if one cannot 
accommodate such principles into the framework, the so called 
topological field theories will just remain formal functional integrals.

In the method presented here, one views the moduli space ${\cal N}_F$ of 
self--dual connections as a finite dimensional submanifold of 
the infinite dimensional space  $\cal N$ of gauge orbits.
Correspondingly, integration of forms on ${\cal N}_F$ is reduced to
integration on $\cal N$ by wedging with the formal Poincar\'e dual
$\chi_{{\cal N}_F}$ of ${\cal N}_F$. Needless to say, this procedure
is rather formal and the possibility of its concrete implementation 
is unclear at the present moment. 

The natural question arises about whether our approach for the treatment
of reducible connections and the analysis of the Gribov ambiguity
may be used in the study of ordinary gauge theories where similar 
problems occur. Indeed this can be done upon performing obvious 
modifications: the $s$ cohomologies relevant in topological and  
ordinary gauge field theories are respectively equivariant and 
BRS cohomology and these are essentially different. The 
application of the corresponding method in ordinary gauge theory
would yield the same gauge fixing sector and the bosonic half
the Mathai--Quillen $\bar\tau,\bar\omega$ sector containing the 
equivariant functional $H_\alpha$. It remains to be seen if this is going 
to provide useful insight.
\vskip.6cm
\par\noindent
{\bf Acknowledgements.} We are greatly indebt to R. Stora for providing 
his invaluable experience and relevant literature. We also thank the
referee of the paper for useful suggestions and improvements.
\vskip.6cm
\centerline{\bf REFERENCES}
\def\ref#1{\lbrack #1\rbrack}
\def\NP#1{Nucl.~Phys.~{\bf #1}}
\def\PL#1{Phys.~Lett.~{\bf #1}}

\def\CMP#1{Commun.~Math.~Phys.~{\bf #1}}
\def\JMP#1{J.~Math.~Phys.~{\bf #1}}

\def\PREP#1{Phys.~Rep.~{\bf #1}}

\def\AP#1{Ann.~Phys.~{\bf #1}}

\vskip.4cm
\par\noindent

\item{\ref{1}} E. Witten, {\it Introduction to Cohomological Field Theories},
Lectures at the Trieste Workshop on Topological Methods in Physics,
Trieste, Italy, June 1990, Int. J. Mod. Phys. {\bf A6} (1991) 2775.

\item{\ref{2}} R. Dijkgraaf, {\it Notes on Topological String Theory and 2-D
Quantum Gravity}, 
presented at the Trieste Spring School School in String Theory and Quantum 
Gravity, Trieste, Italy, April 1990, proceedings, M. Green, R. Iengo, 
S Randjbar--Daemi, H. Verlinde editors, 

\item{\ref{3}} R. Dijkgraaf, {\it Introduction to Topological Field Theories},
in Boulder 1992, Recent Directions in Particle Theory, proceedings 689. 

\item{\ref{4}} D. Birmingham, M. Blau, M. Rakowski and G. Thompson,
{\it Topological Field Theories}, \PREP{209} 4 \& 5 (1991) 129.

\item{\ref{5}} M. Blau and G. Thompson, {\it 1992 Lectures on Topological
Gauge Theory and Yang--Mills Theory}, presented at the Trieste Summer 
School in High Energy Physics and Cosmology, Trieste, Italy, June--July
1992, proceedings, E. Gava, K. Narain, S. Randjbar--Daemi,
E. Sezgin and Q. Shafi editors, hep--th/9305120.

\item{\ref{6}} M. Blau and G. Thompson, {\it Lectures on 2--D Gauge Theories;
Topological Aspects and Path Integral Techniques}, presented at the Trieste
Summer School in High Energy Physics and Cosmology, Trieste, Italy, June--July
1993, proceedings, E. Gava, A. Masiero, K. Narain, S. Randjbar--Daemi, 
and Q. Shafi editors, hep--th/9310144.

\item{\ref{7}} M. Blau and G. Thompson, {\it Localization and Diagonalization, 
a Review of Functional Integral Techniques for Low Dimensional Gauge Theories
and Topological Field Theories}, \JMP{36} (1995) 2192.

\item{\ref{8}} G. Thompson, {\it New Results in Topological Gauge Theory
and Abelian Gauge Theory}, presented at Trieste Summer School in 
High Energy Physics and Cosmology, Trieste, Italy, June--July
1995, hep-th/9511038.

\item{\ref{9}} S. Cordes, G. Moore and S. Ramgoolam, {\it Lectures on 2--D 
Yang--Mills Theory, Equivariant Cohomology and Topological Field Theory},
presented at the Trieste Spring School School in String Theory, Gauge Theory 
and Quantum Gravity, Trieste, Italy, April 1994, and to the Les Houches Summer 
School on Fluctuating Geometries in Statistical Mechanics and Field Theory, 
NATO Advanced Study Institute, Les Houches, France August--September 1994, 
Nucl. Phys. Proc. Suppl. {\bf 141} (1995) 184 (part 1), Les Houches 
Proceedings (part 2), hep--th/9411210.

\item{\ref{10}} J. M. F. Labastida, {\it Topological Quantum Field Theory:
a Progress Report}, presented at the 4--th Fall Workshop on Differential 
Geometry and its Applications, Santiago de Compostela, Spain, September
1995, hep--th/9511037

\item{\ref{11}} E. Witten, {\it Quantum Field Theory and the Jones 
Polynomials}, Commun. Math. Phys. {\bf 121} (1989) 351.

\item{\ref{12}} E. Witten, {\it Topological Quantum Field Theory},
\CMP{117} (1988) 353.

\item{\ref{13}} E. Witten, {\it Topological Sigma Model},
\CMP{118} (1988) 411.

\item{\ref{14}} J. M. F. Labastida, M. Pernici and E. Witten, {\it 
Topological Gravity in Two Dimensions}, \NP{B310} (1988) 611.

\item{\ref{15}} L. Baulieu and I. M. Singer, {\it Topological Yang--Mills
Symmetry}, Nucl. Phys. Proc. Suppl. {\bf 5B} (1988) 12.

\item{\ref{16}} L. Baulieu and I. M. Singer, {\it Conformally Invariant 
Gauge Fixed Actions for 2-D Topological Gravity}, \CMP{135} (1991) 253.

\item{\ref{17}}  J. Kalkman, {\it BRST Model for Equivariant Cohomology and 
Representatives for the Equivariant Thom Class}, \CMP{153} (1993) 447.

\item{\ref{18}} S. Ouvry, R. Stora and P. van Baal, {\it On the
Algebraic Characterization of Witten Topological Yang--Mills Theory},
\PL{B220} (1989) 159.

\item{\ref{19}} R. Stora, F. Thuillier and J.-C. Wallet, {\it 
Algebraic Structure of Cohomological Field Theory Models 
and Equivariant Cohomology}, lectures presented at the First Carribean 
School of Mathematics and Theoretical Physics, Saint Francois,
Guadaloupe May--June 1993.

\item{\ref{20}} V. Mathai and D. Quillen, {\it Superconnections, Thom Classes
and Equivariant Differential Forms}, Topology {\bf 25} (1986) 85.

\item{\ref{21}} M. F. Atiyah and L. Jeffrey, {\it Topological Lagrangians 
and Cohomology}, J. Geom. Phys. 7 (1990) 119.

\item{\ref{22}} M. Blau, {\it The Mathai--Quillen Formalism and Topological
Field Theory}, J. Geom. Phys. {\bf 11} (1991) 129.

\item{\ref{23}} R. Stora, {\it Equivariant Cohomology and Topological Field 
Theories}, presented at the International Symposium on BRS Symmetry
on the Occasion of its 20--th Anniversary, Kyoto Japan, September 1995,
ENSLAPP-A-571-95.

\item{\ref{24}} M. Blau and G. Thompson, {\it Topological Field Theories
of Antisymmetric Tensor Fields}, \AP{205} (1991) 130.

\item{\ref{25}} J. S. Park, {\it Zero Modes, Covariant Anomaly Counterparts
and Reducible Connections in Topological Yang--Mills Theory}, preprint
ESENAT--92--08, hep--th/9210151.

\item{\ref{26}} D. Anselmi, {\it Anomalies in Instanton Calculus}, \NP{B439}
(1995) 617.

\item{\ref{27}} L. Baulieu and M. Schaden, {\it Gauge Group TQFT and 
Improved Perturbative Yang--Mills Theory}, preprint UTH--96--01--05,
hep--th/9601039.

\item{\ref{28}} C. M. Becchi, R. Collina and C. Imbimbo, {\it A Functional
and Lagrangian Formulation of Two--Dimensional Topological Gravity},
hep-th/9406096.

\item{\ref{29}} C. M. Becchi and C. Imbimbo, {\it Gribov Horizon, Contact 
Terms and Cech--de Rham Cohomology in 2D Topological Gravity}, 
hep--th/9510003.

\item{\ref{30}} C. M. Becchi and C. Imbimbo, {\it A Lagrangian 
Formulation of Two--Dimensional Topological Gravity and Cech--deRham 
Cohomology}, presented at the International Symposium on BRS Symmetry on the 
Occasion of Its 20th Anniversary, Kyoto, Japan, September 1995,
hep--th/9511156.

\item{\ref{31}} W. Grueb, S. Halperin and R. Vanstone,
{\it Connections, Curvature and Cohomology}, vol. III, Academic Press, 
New York 1973

\item{\ref{32}} S. Donaldson and P. Kronheimer, {\it The Geometry of
Four-Manifolds}, Clarendon Press, Oxford 1990.

\item{\ref{33}} J. H. Horne, {\it Superspace Version of Topological
Theories}, \NP{B318} (1989) 22.

\item{\ref{34}} G. E. Bredon,  {\it Sheaf Theory}, 
McGraw--Hill Book Company, New York 1967.

\bye